\documentclass[twocolumn, trackchanges]{aastex62}

\submitjournal{ApJ}
\usepackage{amsmath}
\usepackage{comment}
\usepackage{cleveref}

\crefformat{section}{\S#2#1#3} % see manual of cleveref, section 8.2.1
\crefformat{subsection}{\S#2#1#3}
\crefformat{subsubsection}{\S#2#1#3}
\shorttitle{The emergence of the Star Formation Main Sequence}
\shortauthors{Rinaldi et al.}

\begin{document}

\title{\bf {The emergence of the Star Formation Main Sequence with redshift unfolded by JWST}}

\newcommand{\gsim}{{\;\raise0.3ex\hbox{$>$\kern-0.75em\raise-1.1ex\hbox{$\sim$}}\;}}

\correspondingauthor{Pierluigi Rinaldi}
\email{prinaldi@arizona.edu}

\author[0000-0002-5104-8245]{P. Rinaldi}
\affiliation{Kapteyn Astronomical Institute, University of Groningen,
P.O. Box 800, 9700AV Groningen,
The Netherlands
}
\affiliation{Steward Observatory, University of Arizona, 933 North Cherry Avenue, Tucson, AZ 85721, USA}

\author[0000-0001-6066-4624]{R. Navarro-Carrera}
\affiliation{Kapteyn Astronomical Institute, University of Groningen,
P.O. Box 800, 9700AV Groningen,
The Netherlands
}

\author[0000-0001-8183-1460]{K. I. Caputi}
\affiliation{Kapteyn Astronomical Institute, University of Groningen,
P.O. Box 800, 9700AV Groningen,
The Netherlands
}
\affiliation{Cosmic Dawn Center (DAWN), Copenhagen, Denmark
}

\author[0000-0000-0000-0000]{E. Iani}
\affiliation{Kapteyn Astronomical Institute, University of Groningen,
P.O. Box 800, 9700AV Groningen,
The Netherlands
}

\author[0000-0002-3005-1349]{G. \"Ostlin}
\affiliation{Department of Astronomy, Stockholm University, Oscar Klein Centre, AlbaNova University Centre, 106 91 Stockholm, Sweden}

\author[0000-0002-9090-4227]{L. Colina}
\affiliation{Centro de Astrobiolog\'{\i}a (CAB), CSIC-INTA, Ctra. de Ajalvir km 4, Torrej\'on de Ardoz, E-28850, Madrid, Spain}
\affiliation{Cosmic Dawn Center (DAWN), Denmark}

\author[0000-0002-8909-8782]{S. Alberts}
\affiliation{Steward Observatory, University of Arizona, 933 North Cherry Avenue, Tucson, AZ 85721, USA}

\author[0000-0002-7093-1877]{J. \'Alvarez-M\'arquez}
\affiliation{Centro de Astrobiolog\'{\i}a (CAB), CSIC-INTA, Ctra. de Ajalvir km 4, Torrej\'on de Ardoz, E-28850, Madrid, Spain}

\author[0000-0002-8053-8040]{M. Annunziatella}
\affiliation{Centro de Astrobiolog\'{\i}a (CAB), CSIC-INTA, Ctra. de Ajalvir km 4, Torrej\'on de Ardoz, E-28850, Madrid, Spain}

\author[0000-0002-3952-8588]{L. Boogaard}
\affiliation{Max-Planck-Institut f\"ur Astronomie, K\"onigstuhl 17, 69117 Heidelberg, Germany}

\author[0000-0001-6820-0015]{L. Costantin}
\affiliation{Centro de Astrobiolog\'{\i}a (CAB), CSIC-INTA, Ctra. de Ajalvir km 4, Torrej\'on de Ardoz, E-28850, Madrid, Spain}

\author[0000-0002-4571-2306]{J. Hjorth}
\affiliation{DARK, Niels Bohr Institute, University of Copenhagen, Jagtvej 128,
2200 Copenhagen, Denmark}

\author[0000-0001-5710-8395]{D. Langeroodi}
\affiliation{DARK, Niels Bohr Institute, University of Copenhagen, Jagtvej 128, 2200 Copenhagen, Denmark}

\author[0000-0003-0470-8754]{J. Melinder}
\affiliation{Department of Astronomy, Stockholm University, Oscar Klein Centre, AlbaNova University Centre, 106 91 Stockholm, Sweden}

\author[0000-0002-3305-9901]{T. Moutard}
\affiliation{Aix Marseille Univ, CNRS, CNES, LAM, Marseille, France}
\affiliation{European Space Agency (ESA), European Space Astronomy Centre (ESAC), Camino Bajo del Castillo s/n, 28692 Villanueva de la
Canada, Madrid, Spain}

\author[0000-0003-4793-7880]{F. Walter}
\affiliation{Max-Planck-Institut f\"ur Astronomie, K\"onigstuhl 17, 69117 Heidelberg, Germany}

\begin{abstract}

We investigate the correlation between stellar mass ($M_{\star}$) and star formation rate (SFR) across the stellar mass range $\log_{10}(\rm M_{\star}/M_{\odot}) \approx 6-11$. We consider almost 50,000 star-forming galaxies at $z\approx3-7$, leveraging data from COSMOS/SMUVS, JADES/GOODS-S, and MIDIS/XDF. This is the first study spanning such a wide $M_{\star}$ range without relying on gravitational lensing effects. We locate our galaxies on the $\mathrm{SFR - M_{\star}}$ plane to assess how the location of galaxies in the star-formation  main sequence (MS) and starburst (SB) region evolves with $M_{\star}$ and redshift. We find that the two star-forming modes tend to converge at $\log_{10}(\rm M_{\star}/M_{\odot}) < 7$, with all galaxies found in the SB mode. However, deeper observations will be instrumental for reaching lower SFRs and $M_{\star}$ to further validate this scenario. By dissecting our galaxy sample in $M_{\star}$ and redshift, we show that the emergence of the star-formation MS is $M_{\star}$ dependent: while in galaxies with  $\log_{10}(\rm M_{\star}/M_{\odot}) > 9$ the MS is already well in place at $z=5-7$, for galaxies with  $\log_{10}(\rm M_{\star}/M_{\odot}) \approx 7-8$ it only becomes significant at  $z<4$.  Overall, our results are in line with previous findings that the SB mode dominates amongst low stellar-mass galaxies. The earlier emergence of the MS for massive galaxies is consistent with galaxy downsizing.

\end{abstract}

.
\keywords{Galaxies: formation, evolution,  high-redshift, star formation, starburst, Epoch of Reionization}

\section{Introduction}
In recent decades, galaxy surveys up to very high redshifts have significantly advanced our understanding of galaxy evolution, constraining galaxy physical properties such as stellar mass ($\mathrm{M_{\star}}$) and star formation rates (SFR) (\citealt{LeFloch_2005, Ellis_2013, Oesch_2014, Bouwens_2015, Stefanon_2019, Bowler_2020, Bhatawdekar_2021, Bouwens_2021}). These quantities have been extensively used to constrain the process of gas conversion into stars, i.e., the stellar mass assembly (\citealt{Casey_2012, Huillier_2012, Bauer_2013, Jackson_2020}). Statistical analysis of extensive galaxy samples has established a correlation between $\mathrm{M_{\star}}$ and SFR in star-forming galaxies, revealing the “Main Sequence (MS) of star-forming galaxies”, and identified a passive cloud of galaxies with negligible star formation activity (\citealt{Brinchmann_2004, Noeske_2007}). These initial works triggered a vast amount of later papers studying galaxy evolution on the $\mathrm{SFR-M_{\star}}$ plane    \citep[e.g., ][]{Peng_2010,  Speagle_2014, Salmon_2015, Santini_2017}.

The existence of a star formation main sequence indicates that similar mechanisms may drive the growth of both low- and high-mass galaxies \citep[][]{Noeske_2007}. The MS galaxies grow continuously over a long time period from smooth gas accretion \citep[e.g., ][]{Almeida_2014}. In particular, the position of a galaxy on the $\mathrm{SFR-M_{\star}}$ plane has been proposed to be strictly correlated with its evolutionary stage \citep[e.g.,][]{Tacchella_2016}, while the intrinsic scatter of the MS suggests some variety in the star formation histories (SFHs) for galaxies of a given stellar mass \citep[e.g.,][]{Matthee_2019}.

Recent works have shown that the normalization of the $\mathrm{SFR-M_{\star}}$ relation increases over cosmic time, particularly at $z \approx 0-3$, reflecting higher gas accretion rate and, therefore,
higher SFR in the past \citep[e.g.,][]{Whitaker_2012, Whitaker_2014, Iyer_2018, Popesso_2023}. Usually, this relation is modeled as a power-law, log$_{10}$(SFR) = $\alpha$log$_{10}$(M$_{*}$) + $\beta$ (see \citealt{Speagle_2014} for a more detailed study).
  
In addition, other works have analyzed the presence of galaxies with significantly enhanced star formation activity at high redshifts, the so-called {\it starbursts} (SB),  on the $\mathrm{SFR-M_{\star}}$ plane. However, to date, there is no absolute consensus in the literature on the definition of a starburst. Some studies simply consider that SB galaxies are those sources with a very high SFR (order of $\rm 10-100\;M_{\odot}\,yr^{-1}$; see \citealt{Muxlow_2006, Heckman_2006}) and, therefore, they are located several $\sigma$ above the MS, where $\sigma$ refers to the observed scatter of the MS relation (e.g., \citealt{Elbaz_2011, Rodighiero_2011, Schreiber_2015, Lee_2017, Orlitova_2020}). One could adopt instead a criterion more in line with the textbook definition of starburst \citep{Heckman_2006}, which refers to the galaxy birthrate parameter: a galaxy is in a SB episode when its ongoing SFR is much higher than its average past SFR. However, constraining the average past SFR of a galaxy is not trivial, as it assumes knowing its star formation history and age.

More recently, an alternative definition of starbursts for high-redshift galaxies has been empirically proposed: starbursts are galaxies with high specific star formation rates (sSFRs), with 
\(\log_{10}(\mathrm{sSFR/yr^{-1}}) > -7.60\) \citep{Caputi_2017, Caputi_2021}. 
The inverse of the sSFR represents the stellar mass doubling time, with the above limit corresponding to values \(< 4 \times 10^7 \, \mathrm{yr}\). This is comparable to the short star formation timescales (\(10^6\)–\(10^7 \, \mathrm{yr}\)) observed in local starburst regions \citep[e.g.,][]{Leitherer_time_2001}. Additionally, these star formation timescales are consistent with the upper limits set by gas depletion time scales (\(10^8\) yr) observed in local starburst galaxies, where intense star formation rapidly converts the available gas into stars \citep[e.g.,][]{Mihos_1994, Knapen_2009}.

In any case, the origin of the SB phenomenon is not completely understood.  Current theories propose that starbursts may be driven by large-scale gravitational instabilities, which are influenced by stellar self-gravity \citep{Inoue_2016, Romeo_2016, Tadaki_2018}. Alternatively, starbursts might result from multiple star formation bursts, often triggered by galaxy mergers  \citep{Lamastra_2013, Calabro_2019}.

Earlier assessments that SB galaxies played a minor role in cosmic star-formation history were based on data from relatively massive galaxies with M$_{\star} \gtrsim 10^{10} \, \rm M_\odot$ up to $z\approx 2$ \citep{Rodighiero_2011, Sargent_2012, Lamastra_2013}. However, this view has now been reconsidered following the advent of deeper datasets that allow for the study of less massive galaxies.

In this regard, \citet{Caputi_2017}, by studying a sample of galaxies with $\rm M_{\star} \gtrsim 10^{9}\, M_{\odot}$ at $z \approx 4-5$, discovered a significant fraction of SB and found that star-forming galaxies displayed a bimodal distribution (SB/MS) on the SFR-$\rm M_\star$ plane. Later on, \citet{Bisigello_2018} analyzed a deep sample of star-forming galaxies at $z \approx 0-3$ and determined that the fraction of SB galaxies increases both with redshift and towards lower stellar masses. Building on these findings, \citet{Rinaldi_2022} further confirmed the existence of the SB/MS bimodality across a wider range of redshifts ($z \approx 3-6.5$), showing that this bimodality does not depend on the adopted SFR tracer (e.g., UV or H$\alpha$). These previous studies indicate that the SB population is much more significant than previously thought, underscoring the importance of studying low-mass galaxies to fully unveil the role of SBs in galaxy evolution.

In this study we make use of ultra-deep JWST data to extend the study of the star-forming galaxy distribution on the $\mathrm{SFR-M_{\star}}$ plane down to very low stellar masses. complement our analysis with the Spitzer Matching survey of the Ultra-VISTA ultra-deep stripes (SMUVS; \citealt{Ashby_2018}) catalog to trace the high-mass end of the SFR$-\rm M_{\star}$ plane, allowing us to cover in total five decades in stellar mass ($\rm \log_{10}(M_{\star}/M_{\odot}) \approx 6-11$). Our main goal is to understand when the star-formation MS appeared in cosmic time for galaxies of different stellar masses.

The structure of the paper is as follows. Section \ref{section2} provides an overview of our sample, comprising over 50,000 star-forming galaxies, and Section \ref{section3} details our methodology. In Section \ref{section4}, we present our findings. Specifically, in Section \ref{subsection1}, we show the $\rm SFR-M_{\star}$ plane across four redshift bins: $z\approx 2.8 -3.2$, $z\approx 3.2 -3.9$, $z\approx 3.9 -5$, and $z\approx 5 -7$ and validate the SB/MS bimodality at these cosmic epochs. Subsequently, we analyze the sSFR distribution. In Section \ref{subsection2}, we study the emergence of the MS as a function of stellar mass. Finally, in Section \ref{subsection3}, we analyze the emergence of the MS as a function of both cosmic time and stellar mass. Section \ref{section5} summarizes our key findings.

Throughout this paper, we consider a cosmology with $H_{0} = 70\; \rm km\;s^{-1}\;Mpc^{-1}$, $\Omega_{M} = 0.3$, and $\Omega_{\Lambda} =0.7$. All magnitudes are total and refer to the AB system \citep{Oke_1983}. A \citet{Chabrier_2003} initial mass function (IMF) is assumed (0.1--100 M$_{\odot}$). We define SB as galaxies with $\rm sSFR >10^{-7.60} \, yr^{-1}$ \citep{Caputi_2017, Caputi_2021}.

\section{Dataset}\label{section2}
In this work, we made use of data from two fields: the JWST Advanced Deep Extragalactic Survey (JADES) in the GOODS-South (GOODS-S) region (67.7 arcmin$^2$) and the COSMOS/SMUVS survey (0.66 deg$^2$).
\subsection{JADES/GOODS-S}

For this field, we considered the complete dataset from JADES/GOODS-S data release 2 (DR2), covering a total of 67.7 arcmin$^{2}$. This dataset encompasses the Hubble eXtreme Deep Field (XDF; \citealt{Koekemoer_2013}), which includes ultra-deep MIRI data at 5.6 $\mu$m from MIRI Deep Imaging Survey (MIDIS; \"Ostlin et al. 2024, in prep.). We summarize filters and depths in Table \ref{tab:depths}.

\begin{deluxetable}{ccc}
\tabletypesize{\footnotesize}
\tablecolumns{3}
\tablewidth{0pt}
\tablecaption{5$\sigma$ Photometric Depth in JADES/GOODS-S\label{tab:depths}}
\tablehead{
\colhead{Instrument} & \colhead{Filter} & \colhead{Depth (nJy)} \\
\multicolumn{2}{c}{} & \colhead{JADES Deep \& Medium}\\
}
\startdata
HST/ACS & F435W & 2.33 / 10.77\\
HST/ACS & F606W & 3.61 / 6.96\\
HST/ACS & F775W & 2.22 / 15.79\\
HST/ACS & F814W & 8.20 / 7.2\\
HST/ACS & F850LP & 4.28 / 17.53\\ \hline
JWST/NIRCam & F090W & 3.55 / 6.26\\
JWST/NIRCam & F115W & 2.93 / 5.44\\
JWST/NIRCam & F150W & 2.89 / 5.53\\
JWST/NIRCam & F182M & 8.04 / 9.53\\
JWST/NIRCam & F200W & 3.01 / 5.27\\
JWST/NIRCam & F210M & 5.83 / 12.11\\
JWST/NIRCam & F277W & 2.17 / 4.24\\
JWST/NIRCam & F335M & 3.64 / 3.81\\
JWST/NIRCam & F356W & 2.46 / 4.07\\
JWST/NIRCam & F410M & 3.23 / 6.43\\
JWST/NIRCam & F430M & 7.84 / 7.31\\
JWST/NIRCam & F444W & 2.79 / 5.11\\
JWST/NIRCam & F460M & 10.71 / 9.61\\
JWST/NIRCam & F480M & 7.98 / 6.51\\
\hline
JWST/MIRI & F560W & 13.18 \\
\enddata
\tablecomments{Table adapted from \citet{Hainline_cosmos_2024}. Fluxes are in nJy and measured within an aperture of $r = 0.2$\arcsec. JADES Deep and Medium values are separated by a slash (see Figure 3 in \citealt{Hainline_cosmos_2024}). We do not report HST/UVIS (F225W, F275W, and F336W). MIRI/F560W refers to XDF only and its depth was measured in an aperture of $r=0.23$\arcsec.}

\end{deluxetable}

\subsubsection{MIRI data}
In this work, we made use of the MIDIS/F560W ultra-deep observations carried out in December 2022 over Hubble XDF. These data will be described in \citet{Oestlin_MIDIS_2024} and consist of $\approx40$~hours on source taken in the HUDF, which allowed us to reach 28.6~mag (5$\sigma$) for point-like sources measured in an r=0\farcs23 circular aperture (the radius being chosen to ensure a $\approx70$\% encircled energy). For a more comprehensive view on the MIRI data reduction, refer to Section 2.1.2 in \citet{Rinaldi_2023}.
However, only a small fraction of the sources in GOODS-S has a detection in F560W from MIDIS at $z\approx3-7$ ($\lesssim2\%$). More in general, MIRI plays a minimal role in this study because NIRCam alone, given the redshift range probed ($z \approx 3 - 7$), traces the rest-frame spectra of nearly $99\%$ of our sample beyond H$\alpha$ (up to $z \approx 6.5$), ensuring robust estimates of stellar properties (e.g., $M_{\star}$). Indeed, it has been shown in the literature that MIRI detections, when combined with sufficient data from HST and NIRCam (i.e., our worst-case scenario ensures at least 13 bands without MIRI), do not significantly impact the estimation of stellar properties through spectral energy distribution (SED) fitting (see, e.g., \citealt{Boogaard_2023, Alberts_high_2024, Iani_2023, Li_epochs_2024, Lyu_primer_2024}).

\subsubsection{NIRCam data}
We also considered the NIRCam imaging taken by the JWST Advanced Deep Extragalactic Survey, JADES \citep{Eisenstein_2023a}, Data Release 2  (DR2; \citealt{Eisenstein_2023b}), which includes observations from the JWST Extragalactic Medium-band Survey (JEMS; \citealt{Williams_2023}) and the First Reionization Epoch Spectroscopically Complete Observations (FRESCO, \citealt{Oesch_2023}). This dataset provides a total of 14 bands from 0.9 to 4.8~$\mu$m (6 at short-wavelength, SW, and 8 at long-wavelength, LW), with 5$\sigma$ depths ranging from 30.5 to 30.9~mag (measured in a 0.2\arcsec\, radius circular aperture). We remark that JADES is one of the deepest NIRCam surveys on the sky, only matched in depth (in some bands) by the MIDIS/NIRCam-parallel project \citep{Perez_Gonzalez_2023} and The Next Generation Deep Extragalactic Exploratory Public Near-Infrared Slitless Survey, NGDEEP \citep{Bagley_2024}, which means that we can have access to very low-mass galaxies that, prior JWST's launch, were accessible only thanks to lensed fields (e.g., \citealt{Santini_2017, Rinaldi_2022}). 

\subsubsection{HST data}
We obtained all the HST images over GOODS-S from the Hubble Legacy Field  (HLF/GOODS-S). 
The HLF/GOODS-S provides 13 HST bands covering a wide range of wavelengths (0.2$-$1.6$\mu$m), from the UV (WFC3/ UVIS F225W, F275W, and F336W filters), optical (ACS/ WFC F435W, F606W, F775W, F814W, and F850LP filters),  to near-infrared (WFC3/IR F098M, F105W, F125W, F140W and F160W filters). In this work, we only made use of the deepest ones (i.e., F435W, F606W, F775W, F814W, F850LP, F105W, F125W, F140W, F160W). 
We refer the reader to \citet{Whitaker_2019} for a more detailed description of these observations\footnote{The HLF/GOODS-S) imaging is available at \url{https://archive.stsci.edu/ prepds/hlf/};}.

\subsection{COSMOS/SMUVS}

As a complement, we leverage the SMUVS catalog compiled by \citet{Deshmukh_2018} and later updated by \citet{van_Mierlo_2022}, encompassing $\approx 300,000$ {\sl Spitzer} sources with 28-band photometry from $U$ band to 4.5~$\mu$m.

Briefly, SMUVS utilized {\sl Spitzer}'s IRAC at 3.6 and 4.5~$\mu$m across $0.66\; \rm deg^2$ of the COSMOS field \citep[][]{Scoville_2007}, integrating $\approx$ 25 h/pointing and achieving 80\% completeness at $\approx 25.5$~mag (\citealt{Deshmukh_2018}). This survey overlaps with the deepest near-IR and optical data from UltraVISTA (\citealt{McCracken_2012}) and Subaru (\citealt{Taniguchi_2007}), respectively.

The catalog adopted in this work is based on the ultra-deep UltraVISTA HKs stacked image (\citealt{Deshmukh_2018}).  The SED fitting was performed using \textsc{LePHARE}, utilizing synthetic templates from \citet{BC_2003} and a reddening law by \citet{Calzetti_2000}, as detailed in \citet{Deshmukh_2018}. In particular, we focused only on star-forming galaxies at $z\approx 3 - 7$.

By design, SMUVS enables the exploration of a very different region of the parameter space in the $\text{SFR}-M_{\star}$ plane, as it spans an area approximately $35\times$ larger than that covered by JADES in GOODS-S (although it is about three magnitudes shallower). This makes SMUVS particularly effective for probing the high-mass end, which smaller fields cannot adequately sample.

\section{Photometry and SED fitting}\label{section3}
In this section, we provide an overview of the photometry and SED fitting performed for our sources in JADES/GOODS-S. A comprehensive explanation of the photometry and SED fitting for sources in JADES/GOODS-S is presented in \citet{Navarro_2024_bursty}. For the SMUVS sources, detailed methodologies are referenced in \citet{Deshmukh_2018} and \citet{van_Mierlo_2022}. 

\subsection{Photometry}

We adopted \textsc{SExtractor} \citep{SExtractor} to perform the source detection and extract the photometry across all HST and JWST images (with a pixel scale of 30 mas for both HST and NIRCam images and 60 mas for MIRI/F560W). We adopted a detection strategy similar to the one adopted in \citet{Rinaldi_2023, Rinaldi_midis_2024} and \citet{Navarro_2023}. The detection has been performed by using a super stack image combining F277W, F335M, F356W, F410M, and F444W. 

We used \textsc{SExtractor} by considering a \texttt{hot-mode} configuration, following \citet{Galametz_2013} which has been proved being well suited for detecting very faint sources. Particularly, we adopted a detection threshold of $2\sigma$ and a minimum number of contiguous pixels of \texttt{9}.

We recovered more than $85\%$ of the JADES DR2 official catalog. Interestingly, when anti-crossmatching with the official JADES DR2 catalog (\citealt{Eisenstein_2023b}), we found a substantial number of real sources (visually inspected) which have not been reported in the official JADES DR2 catalog (due to the different detection and deblending strategies, especially close to very bright and extended sources).

We built up our photometric catalog following the same approach as the one adopted in \citet{Rinaldi_2023, Rinaldi_midis_2024} and \citet{Navarro_2023}. Briefly, we combined aperture-corrected photometry, adopting circular apertures (i.e., \texttt{MAG\_APER}) of 0.5\arcsec \; diameter, and Kron apertures (i.e., \texttt{MAG\_AUTO} -- \citealt{Kron_1980}). We chose a circular-aperture flux over a Kron flux when the sources were fainter than a given magnitude. In this case, as we were dealing with very deep images, we adopted mag$\rm_{lim} = 27$ as our faint limit for the Kron aperture. The aperture correction for JWST images has been obtained by using the software \textsc{WebbPSF} (\citealt{webbpsf}). When \textsc{SExtractor}, for a given source, was not able to recover any flux, we estimated an upper limit ($3\sigma$): we placed random apertures ($0.25$\arcsec \.radius) in a square box ($20$\arcsec $\times$ $20$\arcsec) around each source and measured the local background, assigning \texttt{-1} as an error. For those sources, instead, with no photometric coverage, we simply put \texttt{-99}. All fluxes have been corrected for Galactic extinction by using \textsc{dustmaps} (\citealt{dustmaps}). Finally, for each source, we imposed a minimum error of 0.05 mag to account for photometric calibration uncertainties and error underestimation by \textsc{SExtractor} \citep{Sonnett_2013}. 

We double-checked our photometry with the one from the official JADES DR2 catalog and found good agreement. See \citet{Navarro_2024_bursty} for more details.

\subsection{SED fitting}
Once we built up our photometric catalog, we performed SED fitting for all our sources. We adopted a two-step process. We first derived photometric redshifts by using \textsc{Eazy} \citep{Brammer_2008} and, then, derived the stellar properties with \textsc{LePHARE} \citep{LePhare_2011}. This guarantees that the same SED fitting code to derive stellar properties was used for both samples (JADES/GOODS-S and SMUVS), ensuring consistency in the derived results.

\begin{figure}[ht!]
    \centering
    \includegraphics[width = 0.45\textwidth]{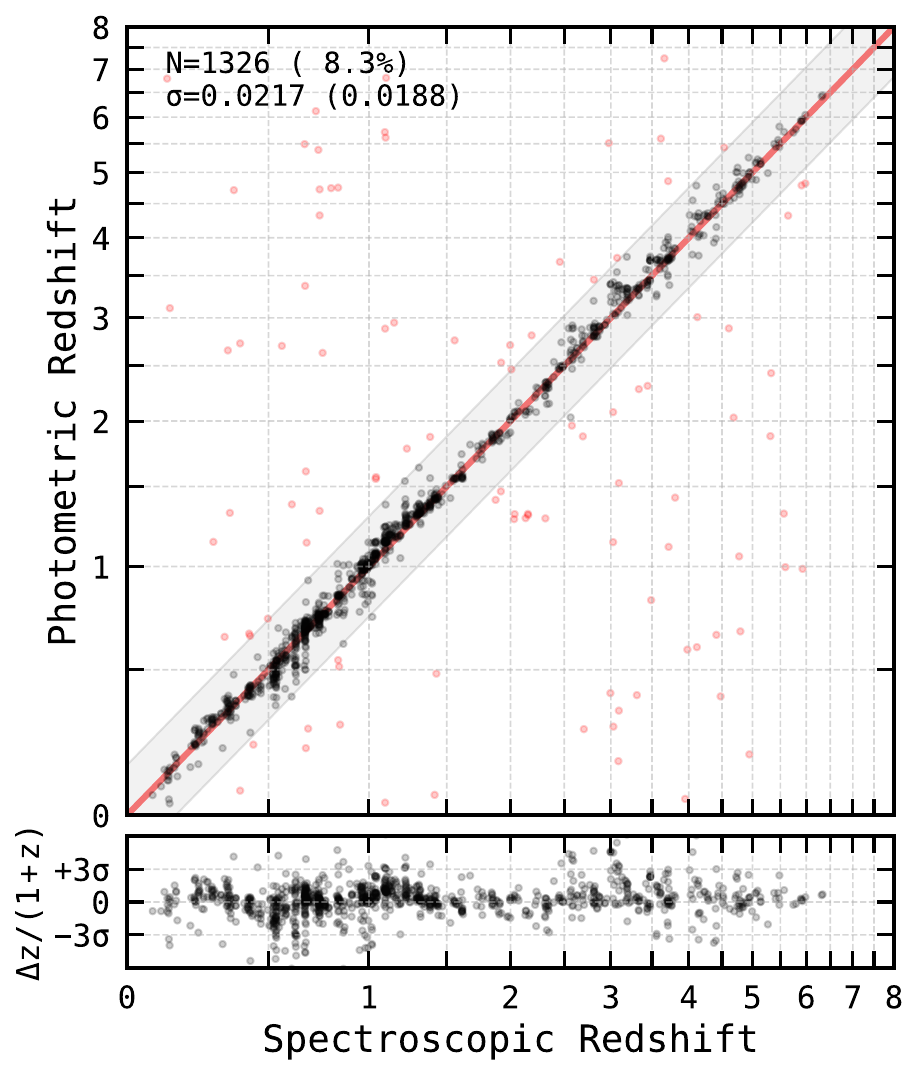}
    \caption{Comparison between spectroscopic and photometric redshifts in JADES/GOODS-S from \citet{Navarro_2024_bursty}.}
    \label{fig:spec-photo-z}
\end{figure}

The \textsc{Eazy} SED fitting has been performed by adopting a linear combination of different templates. Particularly, we adopted the same templates as the ones used in \citet{Eisenstein_2023b, Hainline_cosmos_2024}. This approach allowed us to reach an outlier fraction of $\approx 8\%$ (see Figure \ref{fig:spec-photo-z}, already presented in \citealt{Navarro_2024_bursty}), considering a large amount of spectroscopic redshifts ($>1000$) in GOODS-S from different programs (e.g., JADES/NIRSpec MSA, 3D-HST, MUSE, and CANDELS/GOODS-S; \citealt{Brammer_2012, Guo_2013, Bacon_2023, Bunker_2023}).

Once we derived the photometric redshifts from \textsc{Eazy} and established the goodness of our results, we inferred the stellar properties of our sources with \textsc{LePHARE} by fixing the redshifts to the photometric estimate by \textsc{Eazy} and adopting the same approach as the one used in \citet{Rinaldi_2023, Rinaldi_midis_2024}. Briefly, we considered the stellar population synthesis (SPS) models proposed by \citet{BC_2003}, based on the Chabrier IMF \citep{Chabrier_2003}. We made use of two different star formation histories (SFHs): a standard exponentially declining SFH (known as “$\tau$-model”) and an instantaneous burst adopting a simple stellar population (SSP) model. We opted for two distinct metallicity values, a solar metallicity (Z$_{\odot}$ = 0.02) and a fifth of solar metallicity (Z = 0.2Z$_{\odot}$ = 0.004). We considered the \citet{Calzetti_2000} reddening law in combination with \citet{Leitherer_2002}. In particular, we adopted the following color excess values: $0 \leq E(B-V)\leq 1.5$, with a step of 0.1 mag. {\bf This setup is very similar to the one used by \citet{Deshmukh_2018} and \citet{van_Mierlo_2022} for the SMUVS catalog.}
When a spectroscopic redshift was available, we fixed it to the spectroscopic value rather than the photometric redshift estimated by \textsc{Eazy}.

A more detailed description of the adopted methodology for the SED fitting is presented in \citet{Navarro_2024_bursty}.

\begin{figure*}[ht!]
    \centering
    \includegraphics[width = 0.98 \textwidth, height = 0.60 \textheight]{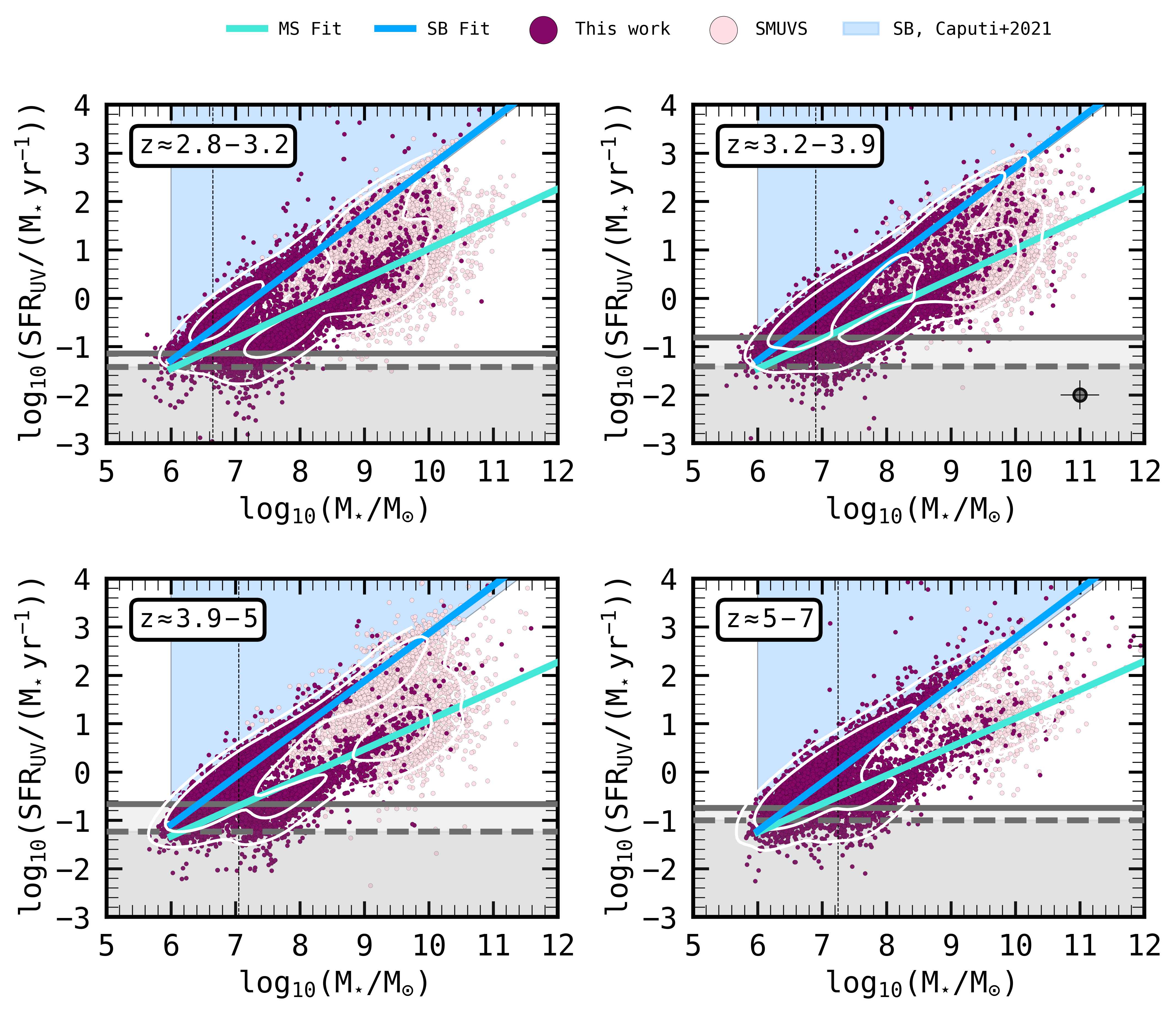}
    \caption{The $\mathrm{SFR}-\mathrm{M_{\star}}$ plane, showcasing all sources (JADES/GOODS-S + COSMOS/SMUVS) analyzed in this study, divided in redshift bins as indicated. The pale blue region marks the lower envelope for SB galaxies, based on the criteria from \citet{Caputi_2017, Caputi_2021}. Fits for the MS and SB are derived from \citet{Rinaldi_2022}. The vertical dashed line in each panel refers to the stellar mass completeness (75\%) of JADES sample at each redshift. The error bar showed in gray (upper right panel) indicate the median uncertainties on $M_{\star}$ and SFR. White contours (68\% and 95\%) are also presented to show the bimodality between MS and SB. The gray horizontal lines indicate the SFR threshold derived from the 3$\sigma$ limiting flux of the band used to estimate the rest-frame 1500 {\AA}. Specifically, when two different filters were used to probe this wavelength for a given redshift bin (e.g., HST/F775W and NIRCam/F090W for $z\approx 3.9-5$), we adopted the average 3$\sigma$ limiting flux. The dashed lines correspond to JADES/Deep, while the solid lines represent JADES/Medium (see Table \ref{tab:depths}).} 
    \label{figure:sfr_m_plane}
\end{figure*}

\section{Results}\label{section4}
We considered our independent determinations of $\rm M_{\star}$ (from SED fitting) and SFRs (from rest-frame UV dust-corrected luminosities) to locate our sample of galaxies (JADES/GOODS-S) on the $\rm SFR - M_{\star}$ plane at $z\approx 3 -7$ complemented with the one from COSMOS/SMUVS, as shown in Figure \ref{figure:sfr_m_plane}. Additionally, we investigated the sSFR distribution of these galaxies, illustrated in Figure \ref{figure:ssfr_m_evo} and Figure \ref{figure:ssfr_z_evo}. Finally, we inspected the evolution of the MS and SB percentages as a function of redshift at different stellar mass regimes (Figure \ref{figure:fraction_evo}).  Our results do not change if SFRs from best-fit SEDs are adopted as SFRs derived from the best-fit models are similarly influenced by dust corrections, as they also rely on the modeled rest-frame UV luminosity.

\subsection{Star Formation Rate estimates from the UV}
We derived the SFRs for our sample independently of their SED fitting by considering their rest-frame UV luminosities ($L_{\nu}$). To do so, we estimated $L_{\nu}$ at a reference wavelength $\lambda_{rest}$ = 1500~{\AA} from the photometry of every galaxy at the filter with the closest effective wavelength to $\lambda_{obs} = \lambda_{rest}  \times (1+z)$, where $z$ is the photometric redshift of that galaxy. We corrected the UV fluxes for dust extinction following the \citet{Calzetti_2000} reddening law in order to recover the intrinsic UV fluxes. In order to do so, we adopted $E(B-V)$ values from the SED fitting analysis. Then we converted them into a monochromatic luminosity ($L_{\nu}$). Finally we obtained SFR$_{UV}$ using the prescription given by \cite{Kennicut_1998}:
\begin{equation}
     \mathrm{SFR (M_{\odot}\;yr^{-1})} = 1.4\times 10^{-28} \; \mathrm{L_{\nu}} \mathrm{(erg} \; \mathrm{s ^{-1} Hz^{-1}).}
     \label{K98}
\end{equation}
The conversion formula (Eq. \ref{K98}) has an intrinsic scatter of 0.3~dex. Therefore, we propagated that error into our uncertainty on the SFR. Furthermore, Eq. \ref{K98} is based on a Salpeter IMF \citep[][]{Salpeter_1955}, while, in this work, we used a Chabrier one. Therefore, we scaled our SFRs from a Salpeter IMF to a Chabrier by considering a conversion factor (i.e., 1.55 as reported in \citealt{Madau_2014}).

\subsection{The bimodality between MS and SB galaxies}\label{subsection1}
Over recent decades, numerous studies have identified that star-forming galaxies predominantly lie along the so-called “Main-Sequence of star-forming galaxies” \citep{Brinchmann_2004, Elbaz_2007, Speagle_2014, Whitaker_2014, Salmon_2015}. 
Only a subset of these sources was classified as starburst galaxies, ranging from about 2\% \citep{Rodighiero_2011} to about 30\% \citep{Caputi_2017}, depending on the stellar mass cut.

However, these prior studies were limited by their observational depth. In this work, we leveraged deep observations from HST and JWST in GOODS-S, with the latter reaching $5\sigma$ depth of about 30.5 mag (\citealt{Eisenstein_2023b, Hainline_cosmos_2024}). These ultra-deep observations enabled us, for the first time, to explore a region of the parameter space that,  previously, was only accessible through the study of lensed fields (i.e., by exploiting the gravitational lensing effect). The latter, prior JWST's launch, was the only way to probe very low-mass galaxies \citep[e.g., ][]{Santini_2017, Atek_2018, Bhatawdekar_2021, Rinaldi_2022}. Notably, in this study, we expanded our dataset with data from SMUVS \citep{Deshmukh_2018}, which allowed us to extend our study toward the high-mass end of the $\rm SFR - M_{\star}$ plane. Overall, our statistical sample consists of more than $\approx 50,000$ galaxies, giving us the opportunity, for the first time, to study the $\rm SFR - M_{\star}$ plane over five decades in stellar mass across cosmic time ($z\approx 3- 7$) in blank fields. 

Following the criteria established in \citet{Caputi_2017, Caputi_2021}, our analysis reveals that approximately $41\%$ of the galaxies in our sample is located along the MS (i.e., they have $\rm log_{10}(sSFR/yr^{-1}) < -8.05$), while about $48\%$ of the sources is placed within the SB cloud (i.e., they have $\rm log_{10}(sSFR/yr^{-1}) > -7.60$). The residual $11\%$ of our sample is located in the so-called “Star Formation Valley” (SFV). The SFV region in the $\rm SFR - M_{\star}$ plane is indicative of galaxies either transitioning from the MS to SB, experiencing a rejuvenation effect \citep{Rosani_2020, Zhang_2023}, or moving from SB back to MS, thereby returning to a steady state after a burst of intense star formation, probably triggered by disk instabilities or (minor and/or major) mergers.

Interestingly, when we apply the same mass cut  (log$\mathbf{_{10}(M_{\star}/M_{\odot}) \gtrsim 10}$) to galaxies as used by \citet{Rodighiero_2011}, we find that SB galaxies constitute only $2\%$ at $z \gtrsim 2.8$. This is in agreement with the findings reported by \citet{Rodighiero_2011} at $z \approx 2$, regardless of which sSFR cut is adopted.

All in all, this comparison suggests that focusing solely on massive galaxies (i.e., $\rm log_{10}(M_{\star}/M_{\odot}) \gtrsim 10$) might lead to a significant underestimation of the role of SB galaxies in the Cosmic Star Formation History, since the starburst phase in high mass galaxies is typically triggered by significant events such as major mergers, as highlighted by studies like \citet{Pearson_2019} and \citet{Renaud_2022}.

Notably, so far, the SB cloud has been defined only down to $\rm log_{10}(M_{\star}/M_{\odot}) \approx 9$ \citep{Caputi_2017, Caputi_2021}. In Figure \ref{figure:sfr_m_plane}, we report an extrapolation toward lower stellar masses ($\rm log_{10}(M_{\star}/M_{\odot}) = 6$) of the SB lower envelope (pale blue shade). Interestingly, from Figure \ref{figure:sfr_m_plane}, we can observe that this extrapolation nicely follows the separation in two clouds between MS and SB galaxies. Indeed, one should note that this SB lower envelope corresponds to the $M_{\star}$ doubling time of approximately \(\rm 4 \times 10^{7}\;yr\), aligning with the typical timescales of local SB episodes as proposed in \citet{Knapen_2009}. Therefore, classifying all galaxies with $\rm log_{10}(sSFR/yr^{-1}) > -7.60$ as SB galaxies is consistent across all $M_{\star}$ ranges, as demonstrated in Figure \ref{figure:sfr_m_plane}.

We then divided our galaxy sample into four distinct redshift bins, following the same approach as the one used in \cite{Rinaldi_2022}. Specifically, \(27\%\) of our sample is located within the redshift range \(z \approx 2.8-3.2\), \(32\%\) falls in the \(z \approx 3.2-3.9\) range, \(24\%\) are in the \(z \approx 3.9 - 5\) interval, and the remaining \(17\%\) is located within \(z \approx 5-7\) range. The above division in redshift bins has been adopted in order to probe the same amount of time ($\approx 400$ Myr) in each redshift bin. Notably, we retrieve the bimodality between MS and SB in each redshift bin. This finding further supports the existence of this bimodality, as previously suggested in \citet{Caputi_2017, Caputi_2021, Rinaldi_2022}. 

As a sanity check, we also inspected all sources in our sample that show a photometric excess ascribable to the H$\alpha$ emission line by using the same approach as the one adopted in \cite{Rinaldi_2023, Caputi_2024}. We first estimated their SFRs(H$\alpha$) and, then, their corresponding sSFR(H$\alpha$). Finally, we looked at the sSFR distribution, confirming that the SB/MS bimodality is evident and does not depend on which SFR is adopted, as already shown in \citet{Rinaldi_2022}.  A complete analysis of the H$\alpha$ emission of those sources is presented in \citet{Navarro_2024_bursty}.

Finally, in Figure \ref{figure:sfr_m_plane}, we note the presence of gray horizontal lines representing the $3\sigma$ limiting fluxes of the bands used to estimate SFR$_{UV}$. Specifically, the dashed lines correspond to the JADES/Deep region of GOODS-S, while the solid lines represent JADES/Medium (see Figure 3 in \citealt{Hainline_cosmos_2024}). Notably, with some scatter, our sample consistently lies above the SFR$_{UV}$ threshold achievable in the JADES/Deep region of GOODS-S across all redshift bins analyzed in this study. The vertical dashed line indicates the $M_{\star}$ completeness limit, corresponding to $75\%$ completeness.

Interestingly, below $\rm log_{10}(M_{\star}/M_{\odot}) < 7$, there is no clear distinction between MS and SB galaxies; instead, galaxies appear to mostly populate the SB cloud. This trend could be explained either by an increased burstiness of star formation that becomes important as we progressively go to lower stellar masses (\citealt{Atek_2022, Navarro_2024_bursty}), or by the current limitations in observational depth. Future deeper observations will be instrumental to explore lower SFRs and $M_{\star}$ to further validate this finding.

\begin{figure*}[ht!]
    \centering
    \includegraphics[width = 0.98 \textwidth, height = 0.65 \textheight]{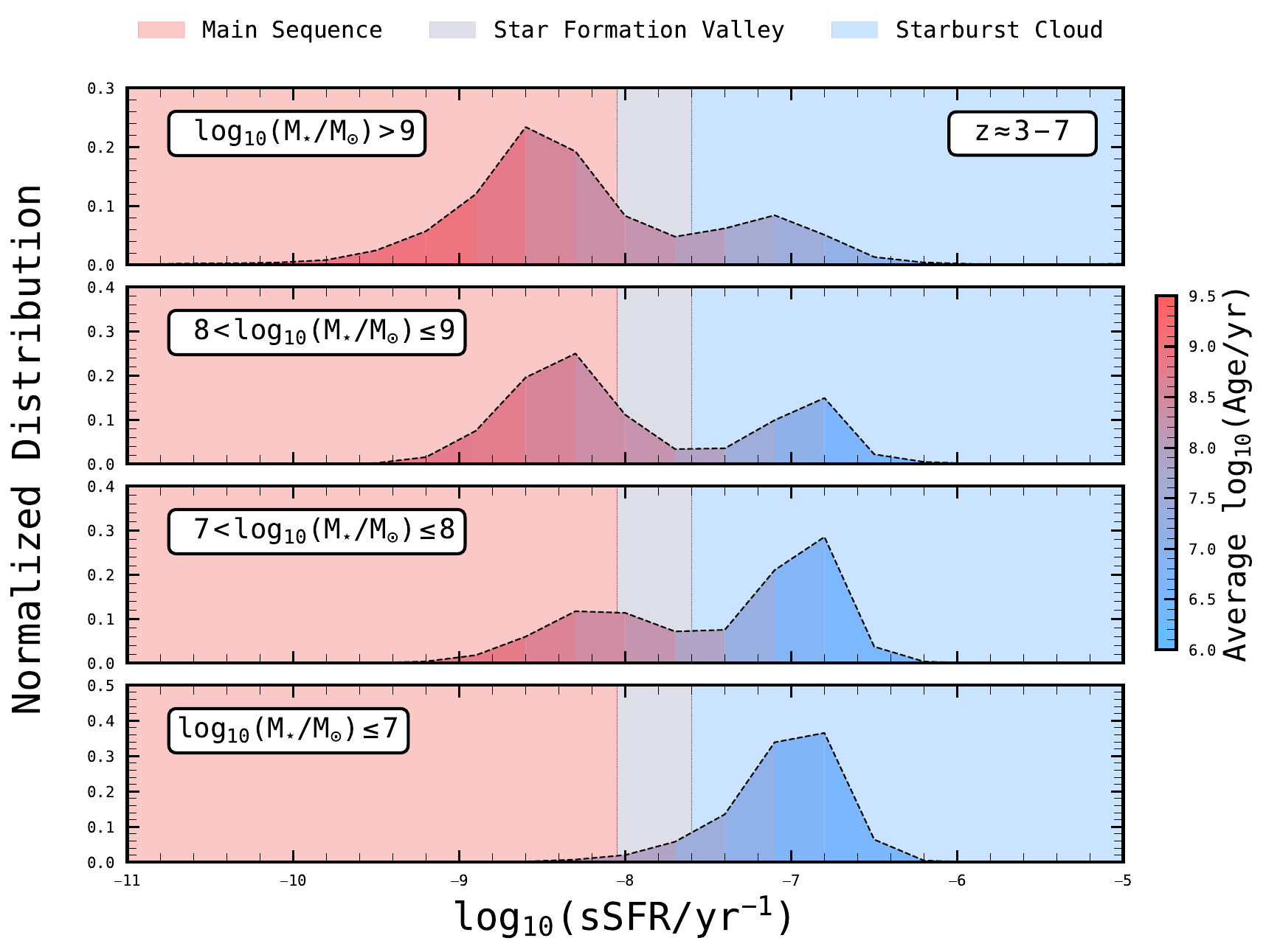}
    \caption{The sSFR distribution of the entire sample (JADES/GOODS-S + COSMOS/SMUVS) divided in four distinct stellar mass bins. The entire plane is colour coded following the regions derived by \citet{Caputi_2017}: the star-formation MS for sSFR $>$ 10$^{-8.05}$ yr$^{-1}$, the Starburst cloud for sSFR $>$ 10$^{-7.60}$ yr$^{-1}$, and the Star Formation Valley for 10$^{-8.05}$ yr$^{-1}$ $\leq$ sSFR $\leq$ 10$^{-7.60}$ yr$^{-1}$. The sSFR distribution are color coded by age, as derived by \textsc{LePHARE}. To consider the different areas covered by JADES/GOODS-S (67.7 arcmin$^{2}$) and COSMOS/SMUVS (0.66 deg$^{2}$), we normalized the JADES/GOODS-S counts to match the COSMOS/SMUVS survey area, which is approximately 35 times larger than that of JADES/GOODS-S.}
    \label{figure:ssfr_m_evo}
\end{figure*}

\subsection{The emergence of the MS across the stellar mass range}\label{subsection2}

Since our sample spans five decades in $M_{\star}$ within the \(\rm SFR - M_{\star}\) plane, we investigated the evolution of the SB/MS bimodality as a function of $M_{\star}$ to identify whether there is a specific stellar mass regime where the “Main Sequence of star-forming galaxies” takes place.

We divided our sample into the following stellar mass bins: \(\rm log_{10}(M_{\star}/M_{\odot}) \leq 7\), \(\rm log_{10}(M_{\star}/M_{\odot}) \approx 7 - 8\), \(\rm log_{10}(M_{\star}/M_{\odot}) \approx 8 - 9\), and \(\rm log_{10}(M_{\star}/M_{\odot}) > 9\). 

Specifically, we analyzed the sSFR distribution within each stellar mass bin. To address the significant differences in survey area between our datasets, we normalized the JADES/GOODS-S (67.7 arcmin$^{2}$) counts to match the COSMOS/SMUVS survey area (0.66 deg$^{2}$), which is approximately 35 times larger than that covered by JADES/GOODS-S.

From Figure \ref{figure:ssfr_m_evo}, we note that in the lowest mass regime (\(\rm log_{10}(M_{\star}/M_{\odot}) \leq 7\)), MS galaxies are almost absent, constituting less than \(3\%\) of the total population at those stellar masses. Interestingly, we find the same result when looking at the sample of H$\alpha$-emitters at $z\gtrsim2.8$ (\citealt{Navarro_2024_bursty}).  By looking at their stellar ages\footnote{The ages for our galaxies directly come from \textsc{LePHARE} and they are purely based on the formation time as given by \citet{BC_2003} models (i.e., the models that we used to perform the SED fitting);} (as indicated by the color bar on the right in Figure \ref{figure:ssfr_m_evo}), we note that MS galaxies, albeit rare, preferentially show an older population compared to the starbursts in the same stellar mass bin. This suggests that MS galaxies may have already reached a steady state (e.g., \citealt{Wang_2019}), with no significant feedback mechanisms triggering new star formation episodes (e.g., \citealt{Renaud_2019}) and, therefore, young and massive stars are not entirely dominating their light. 

Lastly, by looking at the HAEs that fall in this stellar mass regime, we find that galaxies show a very high H$\alpha$ equivalent width (EW) and  are preferentially located in the starburst cloud, which confirms that these sources are going trough a violent episode of star formation. On the contrary, the low-mass HAEs that show a low EW(H$\alpha$) preferentially lie along the MS and represent a very minor fraction of the entire population of HAEs ($\lesssim 1\%$) (\citealt{Navarro_2024_bursty}).

However, we caution the reader that the fact that galaxies preferentially appear to be in a SB phase at lower stellar masses (\(\rm log_{10}(M_{\star}/M_{\odot}) \leq 7\)) could be explained either by an increase in the burstiness of star formation, which becomes more significant as $M_{\star}$ decreases (\citealt{Atek_2022, Navarro_2024_bursty}), or by the limitations of current observational depth which prevent us to detect low-stellar mass objects along the MS. Deeper observations will crucial to shed light on this result.

Moving to higher stellar mass bins, MS galaxies (i.e., those sources with $\text{log}_{10}\text{(sSFR/yr)} < -8.05$) begin to emerge marking the onset of the “Main Sequence of star-forming galaxies”, as we observe for galaxies with stellar mass \(\rm log_{10}(M_{\star}/M_{\odot}) \approx 7 - 8\), and the SB/MS bimodality becomes more prominent, especially at \(\rm log_{10}(M_{\star}/M_{\odot}) \approx 8 - 9\).

As we move to the highest stellar mass bin ($\rm M_{\star} \gtrsim 10^{9}\, M_{\odot}$), MS galaxies dominate the entire sample, now accounting for approximately \(64\%\) at those stellar masses, which favours the picture pointed out in \citet{Rodighiero_2011} where SB galaxies become increasingly rare as we move toward higher $M_{\star}$ (especially at \(\rm log_{10}(M_{\star}/M_{\odot}) \gtrsim 10\)). On the other hand, the HAEs that fall in this stellar mass regime are characterized by having a very low EW(H$\alpha$), due to the strong anti-correlation between $\rm M_{\star}$ and EW(H$\alpha$) (already pointed out in \citealt{Atek_2022} and further demonstrated in \citealt{Navarro_2024_bursty}), indicating a lack of intense star formation activity (\citealt{Navarro_2024_bursty}). The latter may explain why SB galaxies are less common at  \(\rm log_{10}(M_{\star}/M_{\odot}) > 9\), now comprising only  \(26\%\) of the population, with the remaining $\approx 10\%$ in the SFV region. Surprisingly, the fraction of SB galaxies primarily accumulates around  \(\rm log_{10}(M_{\star}/M_{\odot}) \approx 9\) and strongly diminishes at higher stellar masses. 

This result, indeed, indicates that the starburst phenomenon, at very high stellar masses (e.g.,  $\rm M_{\star} \gtrsim 10^{10}\, M_{\odot}$), likely requires galaxies to undergo either a major merger event (although the merger fraction is still not well constrained at high redshift), which could trigger a violent episode of star formation activity (e.g., \citealt{Cibinel_2019, Renaud_2022}), or violent disk instabilities, gas accretion, and/or interactions (flybys and minor mergers) (e.g., \citealt{Bournaud_2012, Dannerbauer_2017, Ho_2019, Pan_2019}).

Interestingly, the results are independent on the assumed SFH. In addition to the initial assumptions discussed in \S 3.2, we explored also a delayed SFH with \textsc{LePHARE}. Despite a different SFH, the trends illustrated in Figure \ref{figure:ssfr_m_evo} remain overall unchanged. Finally, by following \citet{Iani_2023} (see their Appendix C), we investigated whether our results are influenced by the introduction of a correction factor (\textit{k}) to the UV-derived SFR adopted in this study. Specifically, we analyzed the theoretical evolution of the SFR(H$\alpha$)/SFR(UV) ratio as a function of time on a log-log scale, considering the prescriptions of \citet{Kennicut_1998}. This was done for the two metallicities used in our work, namely solar and sub-solar. We employed the same SFH models adopted for the SED fitting of our sources, including single burst and $\tau$-models with $\tau$ values of 0.001, 0.01, 0.03, 1, 2, 3, 5, 8, 10, and 15 Gyr, along with a constant SFH model. These models incorporate the assumptions from the BPASS models \citep{Eldridge_2022}, tailored for a Chabrier IMF, a cutoff mass of 100 $\rm M_{\odot}$, and excluding binary stars. By correcting the SFRs for galaxies younger than 15 Myr—where the UV estimates of SFR are known to be significantly underestimated (e.g., \citealt{Kennicut_1998, Calzetti_2013})—we observed that the overall sSFR distribution remains unaffected.

\subsection{The emergence of the MS over cosmic time}\label{subsection3}

\begin{figure*}[ht!]
    \centering
    \includegraphics[width = 1 \textwidth, height = 0.6 \textheight]{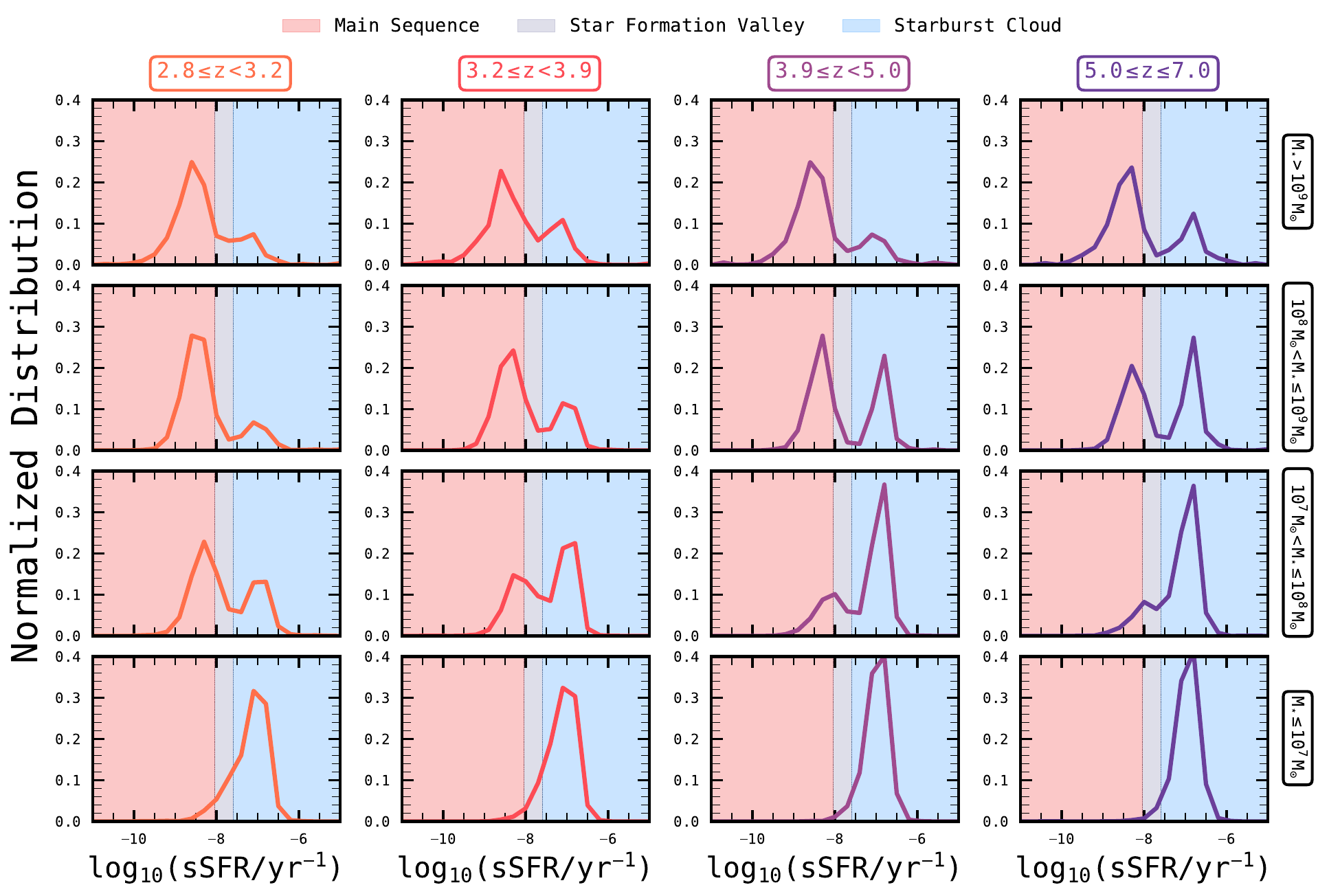}
    \caption{The sSFR distribution of the entire sample (JADES/GOODS-S + COSMOS/SMUVS) divided, this time, in four distinct stellar mass bins and four redshift bins. Each column refers to a specific redshift bin, while each row refers to a specific stellar mass bin. All 16 panels are color coded following \citet{Caputi_2017}: the star-formation MS for sSFR $>$ 10$^{-8.05}$ yr$^{-1}$, the Starburst cloud for sSFR $>$ 10$^{-7.60}$ yr$^{-1}$, and the Star Formation Valley for 10$^{-8.05}$ yr$^{-1}$ $\leq$ sSFR $\leq$ 10$^{-7.60}$ yr$^{-1}$. Also in this case, as we did in Figure \ref{figure:ssfr_m_evo}, to consider the different areas covered by JADES/GOODS-S (67.7 arcmin$^{2}$) and COSMOS/SMUVS (0.66 deg$^{2}$), we normalized the JADES/GOODS-S counts to match the COSMOS/SMUVS survey area, which is approximately 35 times larger than that of JADES/GOODS-S.}
    \label{figure:ssfr_z_evo}
\end{figure*}

We then inspected the emergence of the MS galaxies as a function of cosmic time. For this purpose, as we already assumed in Figure \ref{figure:sfr_m_plane}, we divided our sample into four different redshift bins, each spanning approximately \( \rm 400\; Myr \): \( z \approx 2.8 - 3.2 \), \( z \approx 3.2 - 3.9 \), \( z \approx 3.9 - 5 \), and \( z \approx 5 - 7 \). 

In Figure \ref{figure:ssfr_z_evo}, we illustrate the sSFR distribution as a function of redshift, segmented into four stellar mass bins: we show 16 panels in total, with each row corresponding to a specific stellar mass bin and each column to a specific redshift bin.

As evident from Figure \ref{figure:ssfr_z_evo}, there is a clear evolution of the SB/MS bimodality as a function of cosmic time. Going from the highest redshift bin (\( z \approx 5 - 7 \)) to the lowest one (\( z \approx 2.8 - 3.2 \)), the SB/MS bimodality evolves remarkably, showing that the onset of the MS over cosmic time clearly depends on the stellar mass regime considered.

\begin{figure*}[ht!]
    \centering
    \includegraphics[width = 0.98 \textwidth, height = 0.65 \textheight]{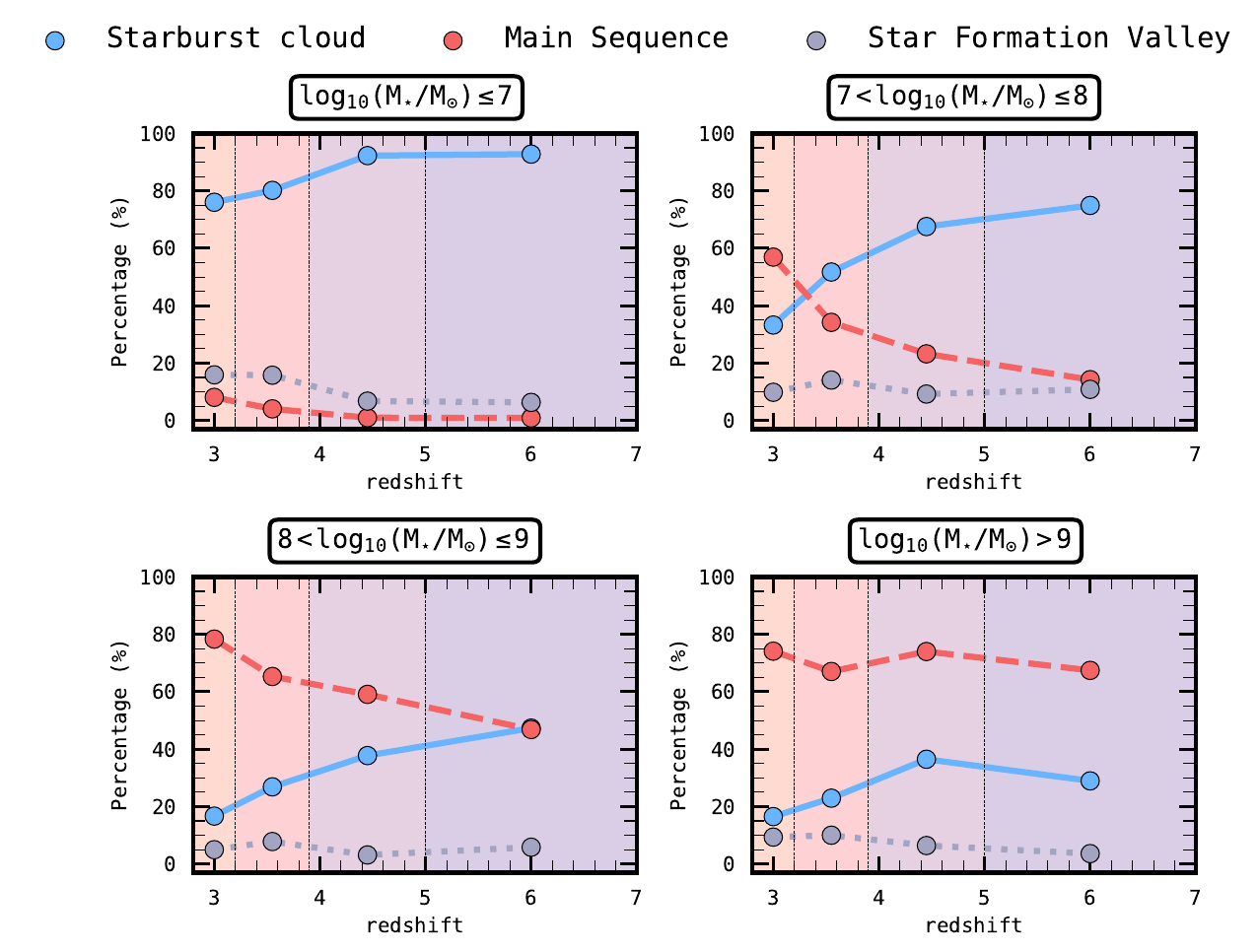}
    \caption{The evolution of the MS, SB, and SFV percentages with cosmic time in four  stellar mass bins. Each panel has color bands corresponding to the four redshift bins analyzed in this study Also in this case, to consider the different areas covered by JADES/GOODS-S (67.7 arcmin$^{2}$) and COSMOS/SMUVS (0.66 deg$^{2}$), we normalized the JADES/GOODS-S counts to match the COSMOS/SMUVS survey area, which is approximately 35 times larger than that of JADES/GOODS-S. }
    \label{figure:fraction_evo}
\end{figure*}

Interestingly, as we move toward lower redshifts, SB galaxies become increasingly less common. Concurrently, we observe the onset of the “Main Sequence of star-forming galaxies”: for galaxies \(\rm M_{\star}\gtrsim10^{8}\, M_{\odot}\), the MS is already in place at \(z\approx 5-7\) (more prominently for \(\rm M_{\star}\gtrsim10^{9}\, M_{\odot}\)).  Notably, MS galaxies are nearly absent in the lowest stellar mass regime, a trend consistent across all redshift bins. 

This evolution over cosmic time suggests that at lower redshifts, the SB phenomenon becomes less common as galaxies have already assembled most of their $M_{\star}$. Significant events, such as major mergers, are thus required to trigger intense star formation episodes in massive galaxies. However, such events become increasingly rare as we move toward lower redshifts ($z\lesssim3$; e.g., \citealt{Ventou_2019}, Pusk\'{a}s et al. in prep.).

This result is also in agreement with what has been presented in \citet{Bisigello_2018}, where they clearly observe that at $z\lesssim3$ (see their Figure 8) the dominant mode of forming stars is the one that regulates the galaxies along the Main-Sequence, while SB galaxies are almost absent.  Finally, in Figure \ref{figure:ssfr_z_evo},  we also observe that within a fixed redshift bin, the SB/MS bimodality evolves as a function of stellar mass, as already shown in Figure \ref{figure:ssfr_m_evo}. 

The emergence of the “Main Sequence of star-forming galaxies” can be further observed in Figure \ref{figure:fraction_evo}, where we show the percentage evolution of MS, SB, and SFV galaxies across redshift and stellar mass bins. To estimate uncertainties in the percentages of MS, SB, and SFV galaxies, we employed Monte Carlo (MC) simulations following the approach used in \citet{Rinaldi_2022}. We generated 1000 mock catalogs from our initial dataset (JADES/GOODS-S + COSMOS/SMUVS), perturbing $M_{\star}$ and SFR within their error bars for each simulation run. Subsequently, these mock galaxies were divided according to the same four stellar mass and redshift bins analyzed in Figure \ref{figure:fraction_evo}, adopting the criteria for MS, SB, and SFV classifications. Then, for each bin, we built up a distribution of the percentages for MS, SB, and SFV galaxies. We then determined the 1$\sigma$ uncertainty as the half-distance between the 16th and 84th percentiles of each distribution. The errors were found to be up to a maximum of 5-6\%.

Consistent with our analysis in Figures \ref{figure:ssfr_m_evo} and Figure \ref{figure:ssfr_z_evo}, MS galaxies are almost absent at \( \rm \log_{10}(M_{\star}/M_{\odot}) \leq 7\). This may be due to either to the fact that very low-mass galaxies generally build up their $\rm M_{\star}$ through more violent episodes of star formation (e.g., \citealt{Asada_2024}), or due to observational constraints that prevent the detection of such low-mass MS galaxies with correspondingly low SFR (below the threshold imposed by the current depth of our observations).

As we consider higher stellar masses (\( \rm \log_{10}(M_{\star}/M_{\odot}) > 7\)), the Main Sequence starts taking place, becoming the dominant mode of forming stars at \( \rm \log_{10}(M_{\star}/M_{\odot}) > 9\) at $z\approx 3-7$. Figure \ref{figure:fraction_evo} highlights again that the evolution of a galaxy along the MS is strictly correlated with $\rm M_{\star}$. The onset of the MS becomes evident well before the Cosmic Noon ($z \approx 2$), particularly for galaxies with $\rm M_{\star} > 10^{9}\,M_{\odot}$, for which the MS is already in place by $z \approx 5-7$. This result suggests that galaxies can start evolving through secular processes well before the peak of Cosmic Star Formation, as suggested both in the past \citep[e.g.,][]{Smit_2016} and more recently \citep{Langeroodi_2024}. The transition from bursty to secular star formation is especially evident in higher-$M_{\star}$ galaxies, which generally achieve a steady state of star formation at earlier epochs. This emphasizes that as galaxies build up their $\rm M_{\star}$, they increasingly favor secular evolution, because significant enhancements in their star formation activity typically require events like (minor and/or major) mergers or disk instabilities (e.g., \citealt{Bournaud_2012, Dannerbauer_2017,Cibinel_2019, Ho_2019, Montero_2019, Pan_2019, Renaud_2022}). This implies that more massive galaxies are less likely to enter a SB phase without such significant events.

\section{Summary and Conclusions}\label{section5}

In this paper, we investigated the emergence of the “Main Sequence of star-forming galaxies” with $M_{\star}$ and redshift. For this purpose, we analyzed over 50,000 galaxies at $z\approx 3-7$ on the $\rm SFR-M_{\star}$ plane, spanning a wide stellar mass range $\rm log_{10}(M_{}/M_{\odot}) \approx 6-11$. This study has been made possible thanks to the joint analysis of ultra-deep JWST data in GOODS-S (FRESCO, JADES, JEMS, and MIDIS), as well as shallower COSMOS/SMUVS data over an area $\approx 35\times$ larger, which allowed us to populate the high-mass end of the $\rm SFR-M_{\star}$ plane. This represents the first study that allows to reach very low stellar masses (\( \rm \log_{10}(M_{\star}/M_{\odot}) <7\)) in blank fields without relying on gravitational lensing effects. We divided our sample into four redshift bins: \( z \approx 2.8 - 3.2 \), \( z \approx 3.2 - 3.9 \), \( z \approx 3.9 - 5 \), and \( z \approx 5 - 7 \). This division allowed us to probe similar amounts of time in each redshift bin ($\approx 400$ Myr).

Our key findings are summarized as follows:

\begin{itemize}
    \item In agreement with previous results (\citealt{Caputi_2017, Rinaldi_2022}), we find a bimodality on the  $\rm SFR-M_{\star}$ plane, such that the vast majority of star-forming galaxies lie either on the MS or SB cloud. This pattern is clearly observed for all galaxies with $\mathrm{\log_{10}(M_{\star}/M_{\odot}) > 7}$ across all redshifts (Figure \ref{figure:sfr_m_plane}).
    
    \item Instead, at stellar masses $\rm \log_{10}(M_{\star}/M_{\odot}) < 7$, the two star-formation modes appear to converge in the $\rm SFR-M_{\star}$ plane (across all redshift bins), such that all low stellar-mass galaxies lie in the SB zone (Figure \ref{figure:ssfr_m_evo}). Although our sample progressively loses completeness at such low stellar masses, we note that, at $z\approx2.9-3.2$, it is still 50\% complete down to $\rm \log_{10}(M_{\star}/M_{\odot}) \approx 6.3$, and 75\% complete down to $\rm \log_{10}(M_{\star}/M_{\odot}) \approx 6.65$.
    Therefore, {\it this is not merely a selection effect.}   Actually, this is expected on physical grounds. If the star formation is driven by the stochastic collapse of giant molecular clouds, a low $M_{\star}$ galaxy will by nature be bursty. For example, a single minor gas accretion event could lead to the formation of $10^7 \, \rm M_\odot$ in 10 Myr, doubling the stellar mass of a low-mass galaxy.  Instead, for a high $M_{\star}$ galaxy, a coordinated global SF event will be required required to produce a SB \citep[e.g.,][]{Gerola_1980, Ostlin_2001, Atek_2022}. Deeper observations will be fundamental to further validate this finding.

    \item As the stellar mass increases, the MS of star-forming galaxies becomes more prominent and dominant, especially for \(\rm log_{10}(M_{\star}/M_{\odot}) > 9\). Notably, at higher stellar masses (\(\rm log_{10}(M_{\star}/M_{\odot}) \gtrsim 8\)), the MS is already established by \(z \gtrsim 4\). More importantly, for galaxies with \(\rm log_{10}(M_{\star}/M_{\odot}) \gtrsim 9\), the MS is already in place at $z\approx 5-7$, aligning with the concept of galaxy downsizing (\citealt{Sparre_2015, Franco_2020}).  In contrast, galaxies with lower stellar masses (\(\rm log_{10}(M_{\star}/M_{\odot}) \lesssim 8\)) at these redshifts ($z\approx 4-7$) are not yet on the MS and remain in the starburst phase (Figure \ref{figure:ssfr_z_evo}). Thus, the emergence of the MS for star-forming galaxies at different redshifts heavily depends on $M_{\star}$ (Figure \ref{figure:fraction_evo}), suggesting that galaxies with higher stellar mass can achieve a steady state well before the Cosmic Noon (\(z \approx 2\); e.g., \citealt{Smit_2016, Langeroodi_2024}).

\end{itemize}

In conclusion, current ultra-deep JWST observations indicate that very low-mass galaxies ($\rm \lesssim 10^{7}\; M_{\odot}$) predominantly experience bursty star formation, with only a few rare cases that can be classified as MS, typically showing signs of an evolved population. However, deeper observations will be crucial to confirm these findings. More importantly, our findings confirm that the emergence of the “Main Sequence of star-forming galaxies” in the $\rm SFR-M_{\star}$ plane is heavily $\rm M_{\star}$ dependent. The emergence of the “Main Sequence of star-forming galaxies” varies across cosmic time, with more massive galaxies ($\rm M_{\star} \gtrsim 10^{9}\,M_{\odot}$) establishing a more regulated, secular evolution already at $z\approx 5-7$. This highlights the critical role of $\rm M_{\star}$ in determining galaxy evolution.

Deeper and wider JWST observations will be instrumental in further constraining the emergence of the MS galaxies within the $\rm SFR-M_{\star}$ plane at different cosmic times.

\acknowledgments

The authors thank an anonymous referee for a careful reading and useful comments on this manuscript. The authors thank Nadav Peleg Brochstein and Daniil Ceban for useful discussion and tests on the effects of different SFHs on the stellar masses.

\vspace{5mm}

This work is based on observations made with the NASA/ESA/CSA JWST. The data were obtained from the Mikulski Archive for Space Telescopes (MAST) at the Space Telescope Science Institute, which is operated by the Association of Universities for Research in Astronomy, Inc., under NASA contract NAS 5-03127 for JWST. These observations are associated with JWST programs GTO \#1180, GO \#1210, GO \#1963, GO \#1895, and \# 3215. The authors acknowledge the FRESCO, JEMS, and \# 3215 teams led by coPIs P. Oesch, C. C. Williams, M. Maseda, D. Eisenstein, and R. Maiolino for developing their observing program with a zero-exclusive-access period. Processing for the JADES NIRCam data release was performed on the lux cluster at the University of California, Santa Cruz, funded by NSF MRI grant AST 1828315. Also based on observations made with the NASA/ESA Hubble Space Telescope obtained from the Space Telescope Science Institute, which is operated by the Association of Universities for Research in Astronomy, Inc., under NASA contract NAS 526555. The data presented in this article were obtained from MAST at the Space Telescope Science Institute. The specific observations analyzed can be accessed via \dataset[DOI: 10.17909/gdyc-7g80, 10.17909/fsc4-dt61, 10.17909/fsc4-dt61, 10.17909/T91019, 10.17909/1rq3-8048, 10.17909/z2gw-mk31]..

\vspace{5mm}
K.I.C. and R.N.C. acknowledge funding from the Dutch Research Council (NWO) through the award of the Vici Grant VI.C.212.036. 
K.I.C. and E.I. acknowledge funding from the Netherlands Research School for Astronomy (NOVA). 
S.A. acknowledges support from the JWST Mid-Infrared Instrument (MIRI) Science Team Lead, grant 80NSSC18K0555, from NASA Goddard Space Flight Center to the University of Arizona.

This work was supported by research grants (VIL16599, VIL54489) from VILLUM FONDEN.

L.C. acknowledges support by grants PIB2021-127718NB-100 and PID2022-139567NB-I00 from the Spanish Ministry of Science and Innovation/State Agency of Research MCIN/AEI/10.13039/501100011033 and by “ERDF A way of making Europe”.

J.A.M., and L.C.,
acknowledge support by grant PIB2021-127718NB-100 from the Spanish  
Ministry of Science
and Innovation/State Agency of Research MCIN/AEI/10.13039/501100011033
and by “ERDF A way of making Europe”.

L.C. thanks the support from the Cosmic Dawn Center received during  
visits to DAWN as international associate.

M.A. acknowledges financial support from Comunidad de
Madrid under Atracci\'on de Talento grant 2020-T2/TIC-19971.

\vspace{5mm}
\facilities{{\sl HST}, {\sl JWST}}.

\software{\textsc{Astropy} \citep{astropy_2018}, 
          \textsc{LePHARE} \citep{LePhare_2011},
          \textsc{NumPy} \citep{Numpy},
          \textsc{pandas} \citep{Pandas}
          \textsc{Photutils} \citep{Photutils}, 
          \textsc{SciPy} \citep{Scipy}
          \textsc{SExtractor} \citep{SExtractor},
          \textsc{TOPCAT} \citep{Topcat}.
          }

\bibliography{References}{}

\begin{thebibliography}{}
\expandafter\ifx\csname natexlab\endcsname\relax\def\natexlab#1{#1}\fi
\providecommand{\url}[1]{\href{#1}{#1}}

\bibitem[{{Alberts} {et~al.}(2024){Alberts}, {Williams}, {Helton}, {Suess}, {Ji}, {Shivaei}, {Lyu}, {Rieke}, {Baker}, {Bonaventura}, {Bunker}, {Carniani}, {Charlot}, {Curtis-Lake}, {D'Eugenio}, {Eisenstein}, {de Graaff}, {Hainline}, {Hausen}, {Johnson}, {Maiolino}, {Parlanti}, {Rieke}, {Robertson}, {Sun}, {Tacchella}, {Willmer}, \& {Willott}}]{Alberts_high_2024}
{Alberts}, S., {Williams}, C.~C., {Helton}, J.~M., {et~al.} 2024, \apj, 975, 85

\bibitem[{{Arnouts} \& {Ilbert}(2011)}]{LePhare_2011}
{Arnouts}, S., \& {Ilbert}, O. 2011, {LePHARE: Photometric Analysis for Redshift Estimate}, , , ascl:1108.009

\bibitem[{{Asada} {et~al.}(2024){Asada}, {Sawicki}, {Abraham}, {Brada{\v{c}}}, {Brammer}, {Desprez}, {Estrada-Carpenter}, {Iyer}, {Martis}, {Matharu}, {Mowla}, {Muzzin}, {Noirot}, {Sarrouh}, {Strait}, {Willott}, \& {Harshan}}]{Asada_2024}
{Asada}, Y., {Sawicki}, M., {Abraham}, R., {et~al.} 2024, \mnras, 527, 11372

\bibitem[{{Ashby} {et~al.}(2018){Ashby}, {Caputi}, {Cowley}, {Deshmukh}, {Dunlop}, {Milvang-Jensen}, {Fynbo}, {Muzzin}, {McCracken}, {Le F{\`e}vre}, {Huang}, \& {Zhang}}]{Ashby_2018}
{Ashby}, M.~L.~N., {Caputi}, K.~I., {Cowley}, W., {et~al.} 2018, \apjs, 237, 39

\bibitem[{{Astropy Collaboration} {et~al.}(2018){Astropy Collaboration}, {Price-Whelan}, {Sip{\H{o}}cz}, {G{\"u}nther}, {Lim}, {Crawford}, {Conseil}, {Shupe}, {Craig}, {Dencheva}, {Ginsburg}, {VanderPlas}, {Bradley}, {P{\'e}rez-Su{\'a}rez}, {de Val-Borro}, {Aldcroft}, {Cruz}, {Robitaille}, {Tollerud}, {Ardelean}, {Babej}, {Bach}, {Bachetti}, {Bakanov}, {Bamford}, {Barentsen}, {Barmby}, {Baumbach}, {Berry}, {Biscani}, {Boquien}, {Bostroem}, {Bouma}, {Brammer}, {Bray}, {Breytenbach}, {Buddelmeijer}, {Burke}, {Calderone}, {Cano Rodr{\'\i}guez}, {Cara}, {Cardoso}, {Cheedella}, {Copin}, {Corrales}, {Crichton}, {D'Avella}, {Deil}, {Depagne}, {Dietrich}, {Donath}, {Droettboom}, {Earl}, {Erben}, {Fabbro}, {Ferreira}, {Finethy}, {Fox}, {Garrison}, {Gibbons}, {Goldstein}, {Gommers}, {Greco}, {Greenfield}, {Groener}, {Grollier}, {Hagen}, {Hirst}, {Homeier}, {Horton}, {Hosseinzadeh}, {Hu}, {Hunkeler}, {Ivezi{\'c}}, {Jain}, {Jenness}, {Kanarek}, {Kendrew}, {Kern}, {Kerzendorf}, {Khvalko}, {King}, {Kirkby}, {Kulkarni},
  {Kumar}, {Lee}, {Lenz}, {Littlefair}, {Ma}, {Macleod}, {Mastropietro}, {McCully}, {Montagnac}, {Morris}, {Mueller}, {Mumford}, {Muna}, {Murphy}, {Nelson}, {Nguyen}, {Ninan}, {N{\"o}the}, {Ogaz}, {Oh}, {Parejko}, {Parley}, {Pascual}, {Patil}, {Patil}, {Plunkett}, {Prochaska}, {Rastogi}, {Reddy Janga}, {Sabater}, {Sakurikar}, {Seifert}, {Sherbert}, {Sherwood-Taylor}, {Shih}, {Sick}, {Silbiger}, {Singanamalla}, {Singer}, {Sladen}, {Sooley}, {Sornarajah}, {Streicher}, {Teuben}, {Thomas}, {Tremblay}, {Turner}, {Terr{\'o}n}, {van Kerkwijk}, {de la Vega}, {Watkins}, {Weaver}, {Whitmore}, {Woillez}, {Zabalza}, \& {Astropy Contributors}}]{astropy_2018}
{Astropy Collaboration}, {Price-Whelan}, A.~M., {Sip{\H{o}}cz}, B.~M., {et~al.} 2018, \aj, 156, 123

\bibitem[{{Atek} {et~al.}(2022){Atek}, {Furtak}, {Oesch}, {van Dokkum}, {Reddy}, {Contini}, {Illingworth}, \& {Wilkins}}]{Atek_2022}
{Atek}, H., {Furtak}, L.~J., {Oesch}, P., {et~al.} 2022, \mnras, 511, 4464

\bibitem[{{Atek} {et~al.}(2018){Atek}, {Richard}, {Kneib}, \& {Schaerer}}]{Atek_2018}
{Atek}, H., {Richard}, J., {Kneib}, J.-P., \& {Schaerer}, D. 2018, \mnras, 479, 5184

\bibitem[{{Bacon} {et~al.}(2023){Bacon}, {Brinchmann}, {Conseil}, {Maseda}, {Nanayakkara}, {Wendt}, {Bacher}, {Mary}, {Weilbacher}, {Krajnovi{\'c}}, {Boogaard}, {Bouch{\'e}}, {Contini}, {Epinat}, {Feltre}, {Guo}, {Herenz}, {Kollatschny}, {Kusakabe}, {Leclercq}, {Michel-Dansac}, {Pello}, {Richard}, {Roth}, {Salvignol}, {Schaye}, {Steinmetz}, {Tresse}, {Urrutia}, {Verhamme}, {Vitte}, {Wisotzki}, \& {Zoutendijk}}]{Bacon_2023}
{Bacon}, R., {Brinchmann}, J., {Conseil}, S., {et~al.} 2023, \aap, 670, A4

\bibitem[{{Bagley} {et~al.}(2024){Bagley}, {Pirzkal}, {Finkelstein}, {Papovich}, {Berg}, {Lotz}, {Leung}, {Ferguson}, {Koekemoer}, {Dickinson}, {Kartaltepe}, {Kocevski}, {Somerville}, {Yung}, {Backhaus}, {Casey}, {Castellano}, {Ch{\'a}vez Ortiz}, {Chworowsky}, {Cox}, {Dav{\'e}}, {Davis}, {Estrada-Carpenter}, {Fontana}, {Fujimoto}, {Gardner}, {Giavalisco}, {Grazian}, {Grogin}, {Hathi}, {Hutchison}, {Jaskot}, {Jung}, {Kewley}, {Kirkpatrick}, {Larson}, {Matharu}, {Natarajan}, {Pentericci}, {P{\'e}rez-Gonz{\'a}lez}, {Ravindranath}, {Rothberg}, {Ryan}, {Shen}, {Simons}, {Snyder}, {Trump}, \& {Wilkins}}]{Bagley_2024}
{Bagley}, M.~B., {Pirzkal}, N., {Finkelstein}, S.~L., {et~al.} 2024, \apjl, 965, L6

\bibitem[{{Bauer} {et~al.}(2013){Bauer}, {Hopkins}, {Gunawardhana}, {Taylor}, {Baldry}, {Bamford}, {Bland-Hawthorn}, {Brough}, {Brown}, {Cluver}, {Colless}, {Conselice}, {Croom}, {Driver}, {Foster}, {Jones}, {Lara-Lopez}, {Liske}, {L{\'o}pez-S{\'a}nchez}, {Loveday}, {Norberg}, {Owers}, {Pimbblet}, {Robotham}, {Sansom}, \& {Sharp}}]{Bauer_2013}
{Bauer}, A.~E., {Hopkins}, A.~M., {Gunawardhana}, M., {et~al.} 2013, \mnras, 434, 209

\bibitem[{{Bertin} \& {Arnouts}(1996)}]{SExtractor}
{Bertin}, E., \& {Arnouts}, S. 1996, \aaps, 117, 393

\bibitem[{{Bhatawdekar} \& {Conselice}(2021)}]{Bhatawdekar_2021}
{Bhatawdekar}, R., \& {Conselice}, C.~J. 2021, \apj, 909, 144

\bibitem[{{Bisigello} {et~al.}(2018){Bisigello}, {Caputi}, {Grogin}, \& {Koekemoer}}]{Bisigello_2018}
{Bisigello}, L., {Caputi}, K.~I., {Grogin}, N., \& {Koekemoer}, A. 2018, \aap, 609, A82

\bibitem[{{Boogaard} {et~al.}(2023){Boogaard}, {Gillman}, {Melinder}, {Walter}, {Colina}, {{\"O}stlin}, {Caputi}, {Iani}, {P{\'e}rez-Gonz{\'a}lez}, {van der Werf}, {Greve}, {Wright}, {Alonso-Herrero}, {{\'A}lvarez-M{\'a}rquez}, {Annunziatella}, {Bik}, {Bosman}, {Costantin}, {Crespo G{\'o}mez}, {Dicken}, {Eckart}, {Hjorth}, {Jermann}, {Labiano}, {Langeroodi}, {Meyer}, {Pei{\ss}ker}, {Pye}, {Rinaldi}, {Tikkanen}, {Topinka}, \& {Henning}}]{Boogaard_2023}
{Boogaard}, L.~A., {Gillman}, S., {Melinder}, J., {et~al.} 2023, arXiv e-prints, arXiv:2308.16895

\bibitem[{{Bournaud} {et~al.}(2012){Bournaud}, {Juneau}, {Le Floc'h}, {Mullaney}, {Daddi}, {Dekel}, {Duc}, {Elbaz}, {Salmi}, \& {Dickinson}}]{Bournaud_2012}
{Bournaud}, F., {Juneau}, S., {Le Floc'h}, E., {et~al.} 2012, \apj, 757, 81

\bibitem[{{Bouwens} {et~al.}(2015){Bouwens}, {Illingworth}, {Oesch}, {Trenti}, {Labb{\'e}}, {Bradley}, {Carollo}, {van Dokkum}, {Gonzalez}, {Holwerda}, {Franx}, {Spitler}, {Smit}, \& {Magee}}]{Bouwens_2015}
{Bouwens}, R.~J., {Illingworth}, G.~D., {Oesch}, P.~A., {et~al.} 2015, \apj, 803, 34

\bibitem[{{Bouwens} {et~al.}(2021){Bouwens}, {Oesch}, {Stefanon}, {Illingworth}, {Labb{\'e}}, {Reddy}, {Atek}, {Montes}, {Naidu}, {Nanayakkara}, {Nelson}, \& {Wilkins}}]{Bouwens_2021}
{Bouwens}, R.~J., {Oesch}, P.~A., {Stefanon}, M., {et~al.} 2021, \aj, 162, 47

\bibitem[{{Bowler} {et~al.}(2020){Bowler}, {Jarvis}, {Dunlop}, {McLure}, {McLeod}, {Adams}, {Milvang-Jensen}, \& {McCracken}}]{Bowler_2020}
{Bowler}, R.~A.~A., {Jarvis}, M.~J., {Dunlop}, J.~S., {et~al.} 2020, \mnras, 493, 2059

\bibitem[{{Bradley} {et~al.}(2021){Bradley}, {Sip{\H{o}}cz}, {Robitaille}, {Tollerud}, {Vin{\'\i}cius}, {Deil}, {Barbary}, {Wilson}, {Busko}, {G{\"u}nther}, {Cara}, {Conseil}, {Bostroem}, {Droettboom}, {Bray}, {Andersen Bratholm}, {Lim}, {Barentsen}, {Craig}, {Rathi}, {Pascual}, {Perren}, {Donath}, {Georgiev}, {De Val-Borro}, {Kerzendorf}, {Bach}, {Quint}, {Souchereau}, \& {Weaver}}]{Photutils}
{Bradley}, L., {Sip{\H{o}}cz}, B., {Robitaille}, T., {et~al.} 2021, {astropy/photutils: 1.0.2}, v1.0.2,  Zenodo, doi:10.5281/zenodo.4453725

\bibitem[{{Brammer} {et~al.}(2008){Brammer}, {van Dokkum}, \& {Coppi}}]{Brammer_2008}
{Brammer}, G.~B., {van Dokkum}, P.~G., \& {Coppi}, P. 2008, \apj, 686, 1503

\bibitem[{{Brammer} {et~al.}(2012){Brammer}, {van Dokkum}, {Franx}, {Fumagalli}, {Patel}, {Rix}, {Skelton}, {Kriek}, {Nelson}, {Schmidt}, {Bezanson}, {da Cunha}, {Erb}, {Fan}, {F{\"o}rster Schreiber}, {Illingworth}, {Labb{\'e}}, {Leja}, {Lundgren}, {Magee}, {Marchesini}, {McCarthy}, {Momcheva}, {Muzzin}, {Quadri}, {Steidel}, {Tal}, {Wake}, {Whitaker}, \& {Williams}}]{Brammer_2012}
{Brammer}, G.~B., {van Dokkum}, P.~G., {Franx}, M., {et~al.} 2012, \apjs, 200, 13

\bibitem[{{Brinchmann} {et~al.}(2004){Brinchmann}, {Charlot}, {White}, {Tremonti}, {Kauffmann}, {Heckman}, \& {Brinkmann}}]{Brinchmann_2004}
{Brinchmann}, J., {Charlot}, S., {White}, S.~D.~M., {et~al.} 2004, \mnras, 351, 1151

\bibitem[{{Bruzual} \& {Charlot}(2003)}]{BC_2003}
{Bruzual}, G., \& {Charlot}, S. 2003, \mnras, 344, 1000

\bibitem[{{Bunker} {et~al.}(2023){Bunker}, {Saxena}, {Cameron}, {Willott}, {Curtis-Lake}, {Jakobsen}, {Carniani}, {Smit}, {Maiolino}, {Witstok}, {Curti}, {D'Eugenio}, {Jones}, {Ferruit}, {Arribas}, {Charlot}, {Chevallard}, {Giardino}, {de Graaff}, {Looser}, {L{\"u}tzgendorf}, {Maseda}, {Rawle}, {Rix}, {Del Pino}, {Alberts}, {Egami}, {Eisenstein}, {Endsley}, {Hainline}, {Hausen}, {Johnson}, {Rieke}, {Rieke}, {Robertson}, {Shivaei}, {Stark}, {Sun}, {Tacchella}, {Tang}, {Williams}, {Willmer}, {Baker}, {Baum}, {Bhatawdekar}, {Bowler}, {Boyett}, {Chen}, {Circosta}, {Helton}, {Ji}, {Kumari}, {Lyu}, {Nelson}, {Parlanti}, {Perna}, {Sandles}, {Scholtz}, {Suess}, {Topping}, {{\"U}bler}, {Wallace}, \& {Whitler}}]{Bunker_2023}
{Bunker}, A.~J., {Saxena}, A., {Cameron}, A.~J., {et~al.} 2023, \aap, 677, A88

\bibitem[{{Calabr{\`o}} {et~al.}(2019){Calabr{\`o}}, {Daddi}, {Fensch}, {Bournaud}, {Cibinel}, {Puglisi}, {Jin}, {Delvecchio}, \& {D'Eugenio}}]{Calabro_2019}
{Calabr{\`o}}, A., {Daddi}, E., {Fensch}, J., {et~al.} 2019, \aap, 632, A98

\bibitem[{{Calzetti}(2013)}]{Calzetti_2013}
{Calzetti}, D. 2013, in Secular Evolution of Galaxies, ed. J.~{Falc{\'o}n-Barroso} \& J.~H. {Knapen}, 419

\bibitem[{{Calzetti} {et~al.}(2000){Calzetti}, {Armus}, {Bohlin}, {Kinney}, {Koornneef}, \& {Storchi-Bergmann}}]{Calzetti_2000}
{Calzetti}, D., {Armus}, L., {Bohlin}, R.~C., {et~al.} 2000, \apj, 533, 682

\bibitem[{{Caputi} {et~al.}(2017){Caputi}, {Deshmukh}, {Ashby}, {Cowley}, {Bisigello}, {Fazio}, {Fynbo}, {Le F{\`e}vre}, {Milvang-Jensen}, \& {Ilbert}}]{Caputi_2017}
{Caputi}, K.~I., {Deshmukh}, S., {Ashby}, M.~L.~N., {et~al.} 2017, \apj, 849, 45

\bibitem[{{Caputi} {et~al.}(2021){Caputi}, {Caminha}, {Fujimoto}, {Kohno}, {Sun}, {Egami}, {Deshmukh}, {Tang}, {Ao}, {Bradley}, {Coe}, {Espada}, {Grillo}, {Hatsukade}, {Knudsen}, {Lee}, {Magdis}, {Morokuma-Matsui}, {Oesch}, {Ouchi}, {Rosati}, {Umehata}, {Valentino}, {Vanzella}, {Wang}, {Wu}, \& {Zitrin}}]{Caputi_2021}
{Caputi}, K.~I., {Caminha}, G.~B., {Fujimoto}, S., {et~al.} 2021, \apj, 908, 146

\bibitem[{{Caputi} {et~al.}(2024){Caputi}, {Rinaldi}, {Iani}, {P{\'e}rez-Gonz{\'a}lez}, {{\"O}stlin}, {Colina}, {Greve}, {N{\o}rgaard-Nielsen}, {Wright}, {{\'A}lvarez-M{\'a}rquez}, {Eckart}, {Hjorth}, {Labiano}, {Le F{\`e}vre}, {Walter}, {van der Werf}, {Boogaard}, {Costantin}, {Crespo G{\'o}mez}, {Gillman}, {Jermann}, {Langeroodi}, {Melinder}, {Peissker}, {G{\"u}del}, {Henning}, {Lagage}, \& {Ray}}]{Caputi_2024}
{Caputi}, K.~I., {Rinaldi}, P., {Iani}, E., {et~al.} 2024, \apj, 969, 159

\bibitem[{{Casey} {et~al.}(2012){Casey}, {Berta}, {B{\'e}thermin}, {Bock}, {Bridge}, {Budynkiewicz}, {Burgarella}, {Chapin}, {Chapman}, {Clements}, {Conley}, {Conselice}, {Cooray}, {Farrah}, {Hatziminaoglou}, {Ivison}, {le Floc'h}, {Lutz}, {Magdis}, {Magnelli}, {Oliver}, {Page}, {Pozzi}, {Rigopoulou}, {Riguccini}, {Roseboom}, {Sanders}, {Scott}, {Seymour}, {Valtchanov}, {Vieira}, {Viero}, \& {Wardlow}}]{Casey_2012}
{Casey}, C.~M., {Berta}, S., {B{\'e}thermin}, M., {et~al.} 2012, \apj, 761, 140

\bibitem[{{Chabrier}(2003)}]{Chabrier_2003}
{Chabrier}, G. 2003, \pasp, 115, 763

\bibitem[{{Cibinel} {et~al.}(2019){Cibinel}, {Daddi}, {Sargent}, {Le Floc'h}, {Liu}, {Bournaud}, {Oesch}, {Amram}, {Calabr{\`o}}, {Duc}, {Pannella}, {Puglisi}, {Perret}, {Elbaz}, \& {Kokorev}}]{Cibinel_2019}
{Cibinel}, A., {Daddi}, E., {Sargent}, M.~T., {et~al.} 2019, \mnras, 485, 5631

\bibitem[{{Dannerbauer} {et~al.}(2017){Dannerbauer}, {Lehnert}, {Emonts}, {Ziegler}, {Altieri}, {De Breuck}, {Hatch}, {Kodama}, {Koyama}, {Kurk}, {Matiz}, {Miley}, {Narayanan}, {Norris}, {Overzier}, {R{\"o}ttgering}, {Sargent}, {Seymour}, {Tanaka}, {Valtchanov}, \& {Wylezalek}}]{Dannerbauer_2017}
{Dannerbauer}, H., {Lehnert}, M.~D., {Emonts}, B., {et~al.} 2017, \aap, 608, A48

\bibitem[{{Deshmukh} {et~al.}(2018){Deshmukh}, {Caputi}, {Ashby}, {Cowley}, {McCracken}, {Fynbo}, {Le F{\`e}vre}, {Milvang-Jensen}, \& {Ilbert}}]{Deshmukh_2018}
{Deshmukh}, S., {Caputi}, K.~I., {Ashby}, M.~L.~N., {et~al.} 2018, \apj, 864, 166

\bibitem[{{Eisenstein} {et~al.}(2023{\natexlab{a}}){Eisenstein}, {Willott}, {Alberts}, {Arribas}, {Bonaventura}, {Bunker}, {Cameron}, {Carniani}, {Charlot}, {Curtis-Lake}, {D'Eugenio}, {Endsley}, {Ferruit}, {Giardino}, {Hainline}, {Hausen}, {Jakobsen}, {Johnson}, {Maiolino}, {Rieke}, {Rieke}, {Rix}, {Robertson}, {Stark}, {Tacchella}, {Williams}, {Willmer}, {Baker}, {Baum}, {Bhatawdekar}, {Boyett}, {Chen}, {Chevallard}, {Circosta}, {Curti}, {Danhaive}, {DeCoursey}, {de Graaff}, {Dressler}, {Egami}, {Helton}, {Hviding}, {Ji}, {Jones}, {Kumari}, {L{\"u}tzgendorf}, {Laseter}, {Looser}, {Lyu}, {Maseda}, {Nelson}, {Parlanti}, {Perna}, {Pusk{\'a}s}, {Rawle}, {Rodr{\'\i}guez Del Pino}, {Sandles}, {Saxena}, {Scholtz}, {Sharpe}, {Shivaei}, {Silcock}, {Simmonds}, {Skarbinski}, {Smit}, {Stone}, {Suess}, {Sun}, {Tang}, {Topping}, {{\"U}bler}, {Villanueva}, {Wallace}, {Whitler}, {Witstok}, \& {Woodrum}}]{Eisenstein_2023a}
{Eisenstein}, D.~J., {Willott}, C., {Alberts}, S., {et~al.} 2023{\natexlab{a}}, arXiv e-prints, arXiv:2306.02465

\bibitem[{{Eisenstein} {et~al.}(2023{\natexlab{b}}){Eisenstein}, {Johnson}, {Robertson}, {Tacchella}, {Hainline}, {Jakobsen}, {Maiolino}, {Bonaventura}, {Bunker}, {Cameron}, {Cargile}, {Curtis-Lake}, {Hausen}, {Pusk{\'a}s}, {Rieke}, {Sun}, {Willmer}, {Willott}, {Alberts}, {Arribas}, {Baker}, {Baum}, {Bhatawdekar}, {Carniani}, {Charlot}, {Chen}, {Chevallard}, {Curti}, {DeCoursey}, {D'Eugenio}, {de Graaff}, {Egami}, {Helton}, {Ji}, {Jones}, {Kumari}, {L{\"u}tzgendorf}, {Laseter}, {Looser}, {Lyu}, {Maseda}, {Nelson}, {Parlanti}, {Rauscher}, {Rawle}, {Rieke}, {Rix}, {Rujopakarn}, {Sandles}, {Saxena}, {Scholtz}, {Sharpe}, {Shivaei}, {Simmonds}, {Smit}, {Topping}, {{\"U}bler}, {Venturi}, {Williams}, {Witstok}, \& {Woodrum}}]{Eisenstein_2023b}
{Eisenstein}, D.~J., {Johnson}, B.~D., {Robertson}, B., {et~al.} 2023{\natexlab{b}}, arXiv e-prints, arXiv:2310.12340

\bibitem[{{Elbaz} {et~al.}(2007){Elbaz}, {Daddi}, {Le Borgne}, {Dickinson}, {Alexander}, {Chary}, {Starck}, {Brandt}, {Kitzbichler}, {MacDonald}, {Nonino}, {Popesso}, {Stern}, \& {Vanzella}}]{Elbaz_2007}
{Elbaz}, D., {Daddi}, E., {Le Borgne}, D., {et~al.} 2007, \aap, 468, 33

\bibitem[{{Elbaz} {et~al.}(2011){Elbaz}, {Dickinson}, {Hwang}, {D{\'\i}az-Santos}, {Magdis}, {Magnelli}, {Le Borgne}, {Galliano}, {Pannella}, {Chanial}, {Armus}, {Charmandaris}, {Daddi}, {Aussel}, {Popesso}, {Kartaltepe}, {Altieri}, {Valtchanov}, {Coia}, {Dannerbauer}, {Dasyra}, {Leiton}, {Mazzarella}, {Alexander}, {Buat}, {Burgarella}, {Chary}, {Gilli}, {Ivison}, {Juneau}, {Le Floc'h}, {Lutz}, {Morrison}, {Mullaney}, {Murphy}, {Pope}, {Scott}, {Brodwin}, {Calzetti}, {Cesarsky}, {Charlot}, {Dole}, {Eisenhardt}, {Ferguson}, {F{\"o}rster Schreiber}, {Frayer}, {Giavalisco}, {Huynh}, {Koekemoer}, {Papovich}, {Reddy}, {Surace}, {Teplitz}, {Yun}, \& {Wilson}}]{Elbaz_2011}
{Elbaz}, D., {Dickinson}, M., {Hwang}, H.~S., {et~al.} 2011, \aap, 533, A119

\bibitem[{{Eldridge} \& {Stanway}(2022)}]{Eldridge_2022}
{Eldridge}, J.~J., \& {Stanway}, E.~R. 2022, \araa, 60, 455

\bibitem[{{Ellis} {et~al.}(2013){Ellis}, {McLure}, {Dunlop}, {Robertson}, {Ono}, {Schenker}, {Koekemoer}, {Bowler}, {Ouchi}, {Rogers}, {Curtis-Lake}, {Schneider}, {Charlot}, {Stark}, {Furlanetto}, \& {Cirasuolo}}]{Ellis_2013}
{Ellis}, R.~S., {McLure}, R.~J., {Dunlop}, J.~S., {et~al.} 2013, \apjl, 763, L7

\bibitem[{{Franco} {et~al.}(2020){Franco}, {Elbaz}, {Zhou}, {Magnelli}, {Schreiber}, {Ciesla}, {Dickinson}, {Nagar}, {Magdis}, {Alexander}, {B{\'e}thermin}, {Demarco}, {Daddi}, {Wang}, {Mullaney}, {Sargent}, {Inami}, {Shu}, {Bournaud}, {Chary}, {Coogan}, {Ferguson}, {Finkelstein}, {Giavalisco}, {G{\'o}mez-Guijarro}, {Iono}, {Juneau}, {Lagache}, {Lin}, {Motohara}, {Okumura}, {Pannella}, {Papovich}, {Pope}, {Rujopakarn}, {Silverman}, \& {Xiao}}]{Franco_2020}
{Franco}, M., {Elbaz}, D., {Zhou}, L., {et~al.} 2020, \aap, 643, A30

\bibitem[{{Galametz} {et~al.}(2013){Galametz}, {Grazian}, {Fontana}, {Ferguson}, {Ashby}, {Barro}, {Castellano}, {Dahlen}, {Donley}, {Faber}, {Grogin}, {Guo}, {Huang}, {Kocevski}, {Koekemoer}, {Lee}, {McGrath}, {Peth}, {Willner}, {Almaini}, {Cooper}, {Cooray}, {Conselice}, {Dickinson}, {Dunlop}, {Fazio}, {Foucaud}, {Gardner}, {Giavalisco}, {Hathi}, {Hartley}, {Koo}, {Lai}, {de Mello}, {McLure}, {Lucas}, {Paris}, {Pentericci}, {Santini}, {Simpson}, {Sommariva}, {Targett}, {Weiner}, {Wuyts}, \& {CANDELS Team}}]{Galametz_2013}
{Galametz}, A., {Grazian}, A., {Fontana}, A., {et~al.} 2013, \apjs, 206, 10

\bibitem[{{Gerola} {et~al.}(1980){Gerola}, {Seiden}, \& {Schulman}}]{Gerola_1980}
{Gerola}, H., {Seiden}, P.~E., \& {Schulman}, L.~S. 1980, \apj, 242, 517

\bibitem[{{Green}(2018)}]{dustmaps}
{Green}, G. 2018, The Journal of Open Source Software, 3, 695

\bibitem[{{Guo} {et~al.}(2013){Guo}, {Ferguson}, {Giavalisco}, {Barro}, {Willner}, {Ashby}, {Dahlen}, {Donley}, {Faber}, {Fontana}, {Galametz}, {Grazian}, {Huang}, {Kocevski}, {Koekemoer}, {Koo}, {McGrath}, {Peth}, {Salvato}, {Wuyts}, {Castellano}, {Cooray}, {Dickinson}, {Dunlop}, {Fazio}, {Gardner}, {Gawiser}, {Grogin}, {Hathi}, {Hsu}, {Lee}, {Lucas}, {Mobasher}, {Nandra}, {Newman}, \& {van der Wel}}]{Guo_2013}
{Guo}, Y., {Ferguson}, H.~C., {Giavalisco}, M., {et~al.} 2013, \apjs, 207, 24

\bibitem[{{Hainline} {et~al.}(2024){Hainline}, {Johnson}, {Robertson}, {Tacchella}, {Helton}, {Sun}, {Eisenstein}, {Simmonds}, {Topping}, {Whitler}, {Willmer}, {Rieke}, {Suess}, {Hviding}, {Cameron}, {Alberts}, {Baker}, {Baum}, {Bhatawdekar}, {Bonaventura}, {Boyett}, {Bunker}, {Carniani}, {Charlot}, {Chevallard}, {Chen}, {Curti}, {Curtis-Lake}, {D'Eugenio}, {Egami}, {Endsley}, {Hausen}, {Ji}, {Looser}, {Lyu}, {Maiolino}, {Nelson}, {Pusk{\'a}s}, {Rawle}, {Sandles}, {Saxena}, {Smit}, {Stark}, {Williams}, {Willott}, \& {Witstok}}]{Hainline_cosmos_2024}
{Hainline}, K.~N., {Johnson}, B.~D., {Robertson}, B., {et~al.} 2024, \apj, 964, 71

\bibitem[{Harris {et~al.}(2020)Harris, Millman, van~der Walt, Gommers, Virtanen, Cournapeau, Wieser, Taylor, Berg, Smith, Kern, Picus, Hoyer, van Kerkwijk, Brett, Haldane, del R{\'{i}}o, Wiebe, Peterson, G{\'{e}}rard-Marchant, Sheppard, Reddy, Weckesser, Abbasi, Gohlke, \& Oliphant}]{Numpy}
Harris, C.~R., Millman, K.~J., van~der Walt, S.~J., {et~al.} 2020, Nature, 585, 357.
\newblock \url{https://doi.org/10.1038/s41586-020-2649-2}

\bibitem[{Heckman(2006)}]{Heckman_2006}
Heckman, T. 2006, in Encyclopedia of Astronomy and Astrophysics (Murdin)

\bibitem[{{Ho} {et~al.}(2019){Ho}, {Martin}, \& {Turner}}]{Ho_2019}
{Ho}, S.~H., {Martin}, C.~L., \& {Turner}, M.~L. 2019, \apj, 875, 54

\bibitem[{{Iani} {et~al.}(2024){Iani}, {Caputi}, {Rinaldi}, {Annunziatella}, {Boogaard}, {{\"O}stlin}, {Costantin}, {Gillman}, {P{\'e}rez-Gonz{\'a}lez}, {Colina}, {Greve}, {Wright}, {Alonso-Herrero}, {{\'A}lvarez-M{\'a}rquez}, {Bik}, {Bosman}, {Crespo G{\'o}mez}, {Eckart}, {Hjorth}, {Jermann}, {Labiano}, {Langeroodi}, {Melinder}, {Moutard}, {Pei{\ss}ker}, {Pye}, {Tikkanen}, {van der Werf}, {Walter}, {Henning}, {Lagage}, \& {van Dishoeck}}]{Iani_2023}
{Iani}, E., {Caputi}, K.~I., {Rinaldi}, P., {et~al.} 2024, \apj, 963, 97

\bibitem[{{Inoue} {et~al.}(2016){Inoue}, {Dekel}, {Mandelker}, {Ceverino}, {Bournaud}, \& {Primack}}]{Inoue_2016}
{Inoue}, S., {Dekel}, A., {Mandelker}, N., {et~al.} 2016, \mnras, 456, 2052

\bibitem[{{Iyer} {et~al.}(2018){Iyer}, {Gawiser}, {Dav{\'e}}, {Davis}, {Finkelstein}, {Kodra}, {Koekemoer}, {Kurczynski}, {Newman}, {Pacifici}, \& {Somerville}}]{Iyer_2018}
{Iyer}, K., {Gawiser}, E., {Dav{\'e}}, R., {et~al.} 2018, \apj, 866, 120

\bibitem[{{Jackson} {et~al.}(2020){Jackson}, {Pasquali}, {Pacifici}, {Engler}, {Pillepich}, \& {Grebel}}]{Jackson_2020}
{Jackson}, T.~M., {Pasquali}, A., {Pacifici}, C., {et~al.} 2020, \mnras, 497, 4262

\bibitem[{{Kennicutt}(1998)}]{Kennicut_1998}
{Kennicutt}, Robert~C., J. 1998, \araa, 36, 189

\bibitem[{{Knapen} \& {James}(2009)}]{Knapen_2009}
{Knapen}, J.~H., \& {James}, P.~A. 2009, \apj, 698, 1437

\bibitem[{{Koekemoer} {et~al.}(2013){Koekemoer}, {Ellis}, {McLure}, {Dunlop}, {Robertson}, {Ono}, {Schenker}, {Ouchi}, {Bowler}, {Rogers}, {Curtis-Lake}, {Schneider}, {Charlot}, {Stark}, {Furlanetto}, {Cirasuolo}, {Wild}, \& {Targett}}]{Koekemoer_2013}
{Koekemoer}, A.~M., {Ellis}, R.~S., {McLure}, R.~J., {et~al.} 2013, \apjs, 209, 3

\bibitem[{{Kron}(1980)}]{Kron_1980}
{Kron}, R.~G. 1980, \apjs, 43, 305

\bibitem[{{Lamastra} {et~al.}(2013){Lamastra}, {Menci}, {Fiore}, \& {Santini}}]{Lamastra_2013}
{Lamastra}, A., {Menci}, N., {Fiore}, F., \& {Santini}, P. 2013, \aap, 552, A44

\bibitem[{{Langeroodi} \& {Hjorth}(2024)}]{Langeroodi_2024}
{Langeroodi}, D., \& {Hjorth}, J. 2024, arXiv e-prints, arXiv:2404.13045

\bibitem[{{Le Floc'h} {et~al.}(2005){Le Floc'h}, {Papovich}, {Dole}, {Bell}, {Lagache}, {Rieke}, {Egami}, {P{\'e}rez-Gonz{\'a}lez}, {Alonso-Herrero}, {Rieke}, {Blaylock}, {Engelbracht}, {Gordon}, {Hines}, {Misselt}, {Morrison}, \& {Mould}}]{LeFloch_2005}
{Le Floc'h}, E., {Papovich}, C., {Dole}, H., {et~al.} 2005, \apj, 632, 169

\bibitem[{{Lee} {et~al.}(2017){Lee}, {Sheth}, {Scott}, {Toft}, {Magdis}, {Damjanov}, {Zahid}, {Casey}, {Cortzen}, {G{\'o}mez Guijarro}, {Karim}, {Leslie}, \& {Schinnerer}}]{Lee_2017}
{Lee}, N., {Sheth}, K., {Scott}, K.~S., {et~al.} 2017, \mnras, 471, 2124

\bibitem[{{Leitherer}(2001)}]{Leitherer_time_2001}
{Leitherer}, C. 2001, in Astronomical Society of the Pacific Conference Series, Vol. 245, Astrophysical Ages and Times Scales, ed. T.~{von Hippel}, C.~{Simpson}, \& N.~{Manset}, 390

\bibitem[{{Leitherer} {et~al.}(2002){Leitherer}, {Li}, {Calzetti}, \& {Heckman}}]{Leitherer_2002}
{Leitherer}, C., {Li}, I.~H., {Calzetti}, D., \& {Heckman}, T.~M. 2002, \apjs, 140, 303

\bibitem[{{L'Huillier} {et~al.}(2012){L'Huillier}, {Combes}, \& {Semelin}}]{Huillier_2012}
{L'Huillier}, B., {Combes}, F., \& {Semelin}, B. 2012, \aap, 544, A68

\bibitem[{{Li} {et~al.}(2024){Li}, {Conselice}, {Adams}, {Trussler}, {Austin}, {Harvey}, {Ferreira}, {Caruana}, {Ormerod}, \& {Juod{\v{z}}balis}}]{Li_epochs_2024}
{Li}, Q., {Conselice}, C.~J., {Adams}, N., {et~al.} 2024, \mnras, 531, 617

\bibitem[{{Lyu} {et~al.}(2024){Lyu}, {Magnelli}, {Elbaz}, {P{\'e}rez-Gonz{\'a}lez}, {Correa}, {Daddi}, {G{\'o}mez-Guijarro}, {Dunlop}, {Grogin}, {Koekemoer}, {McLeod}, \& {Lu}}]{Lyu_primer_2024}
{Lyu}, Y., {Magnelli}, B., {Elbaz}, D., {et~al.} 2024, arXiv e-prints, arXiv:2406.11571

\bibitem[{{Madau} \& {Dickinson}(2014)}]{Madau_2014}
{Madau}, P., \& {Dickinson}, M. 2014, \araa, 52, 415

\bibitem[{{Matthee} \& {Schaye}(2019)}]{Matthee_2019}
{Matthee}, J., \& {Schaye}, J. 2019, \mnras, 484, 915

\bibitem[{{McCracken} {et~al.}(2012){McCracken}, {Milvang-Jensen}, {Dunlop}, {Franx}, {Fynbo}, {Le F{\`e}vre}, {Holt}, {Caputi}, {Goranova}, {Buitrago}, {Emerson}, {Freudling}, {Hudelot}, {L{\'o}pez-Sanjuan}, {Magnard}, {Mellier}, {M{\o}ller}, {Nilsson}, {Sutherland}, {Tasca}, \& {Zabl}}]{McCracken_2012}
{McCracken}, H.~J., {Milvang-Jensen}, B., {Dunlop}, J., {et~al.} 2012, \aap, 544, A156

\bibitem[{{Mihos} \& {Hernquist}(1994)}]{Mihos_1994}
{Mihos}, J.~C., \& {Hernquist}, L. 1994, \apjl, 431, L9

\bibitem[{{Muxlow} {et~al.}(2006){Muxlow}, {Beswick}, {Richards}, \& {Thrall}}]{Muxlow_2006}
{Muxlow}, T., {Beswick}, R.~J., {Richards}, A.~M.~S., \& {Thrall}, H.~J. 2006, in Proceedings of the 8th European VLBI Network Symposium, ed. W.~{Baan}, R.~{Bachiller}, R.~{Booth}, P.~{Charlot}, P.~{Diamond}, M.~{Garrett}, X.~{Hong}, J.~{Jonas}, A.~{Kus}, F.~{Mantovani}, A.~{Marecki}, H.~{Olofsson}, W.~{Schlueter}, M.~{Tornikoski}, N.~{Wang}, \& A.~{Zensus}, 31

\bibitem[{{Navarro-Carrera} {et~al.}(2024{\natexlab{a}}){Navarro-Carrera}, {Rinaldi}, {Caputi}, {Iani}, {Kokorev}, {Kerutt}, \& {Cooper}}]{Navarro_2024_bursty}
{Navarro-Carrera}, R., {Rinaldi}, P., {Caputi}, K.~I., {et~al.} 2024{\natexlab{a}}, arXiv e-prints, arXiv:2410.23249

\bibitem[{{Navarro-Carrera} {et~al.}(2024{\natexlab{b}}){Navarro-Carrera}, {Rinaldi}, {Caputi}, {Iani}, {Kokorev}, \& {van Mierlo}}]{Navarro_2023}
---. 2024{\natexlab{b}}, \apj, 961, 207

\bibitem[{{Noeske} {et~al.}(2007){Noeske}, {Weiner}, {Faber}, {Papovich}, {Koo}, {Somerville}, {Bundy}, {Conselice}, {Newman}, {Schiminovich}, {Le Floc'h}, {Coil}, {Rieke}, {Lotz}, {Primack}, {Barmby}, {Cooper}, {Davis}, {Ellis}, {Fazio}, {Guhathakurta}, {Huang}, {Kassin}, {Martin}, {Phillips}, {Rich}, {Small}, {Willmer}, \& {Wilson}}]{Noeske_2007}
{Noeske}, K.~G., {Weiner}, B.~J., {Faber}, S.~M., {et~al.} 2007, \apjl, 660, L43

\bibitem[{{Oesch} {et~al.}(2014){Oesch}, {Bouwens}, {Illingworth}, {Labb{\'e}}, {Smit}, {Franx}, {van Dokkum}, {Momcheva}, {Ashby}, {Fazio}, {Huang}, {Willner}, {Gonzalez}, {Magee}, {Trenti}, {Brammer}, {Skelton}, \& {Spitler}}]{Oesch_2014}
{Oesch}, P.~A., {Bouwens}, R.~J., {Illingworth}, G.~D., {et~al.} 2014, \apj, 786, 108

\bibitem[{{Oesch} {et~al.}(2023){Oesch}, {Brammer}, {Naidu}, {Bouwens}, {Chisholm}, {Illingworth}, {Matthee}, {Nelson}, {Qin}, {Reddy}, {Shapley}, {Shivaei}, {van Dokkum}, {Weibel}, {Whitaker}, {Wuyts}, {Covelo-Paz}, {Endsley}, {Fudamoto}, {Giovinazzo}, {Herard-Demanche}, {Kerutt}, {Kramarenko}, {Labbe}, {Leonova}, {Lin}, {Magee}, {Marchesini}, {Maseda}, {Mason}, {Matharu}, {Meyer}, {Neufeld}, {Prieto Lyon}, {Schaerer}, {Sharma}, {Shuntov}, {Smit}, {Stefanon}, {Wyithe}, \& {Xiao}}]{Oesch_2023}
{Oesch}, P.~A., {Brammer}, G., {Naidu}, R.~P., {et~al.} 2023, \mnras, 525, 2864

\bibitem[{{Oke} \& {Gunn}(1983)}]{Oke_1983}
{Oke}, J.~B., \& {Gunn}, J.~E. 1983, \apj, 266, 713

\bibitem[{{Orlitova}(2020)}]{Orlitova_2020}
{Orlitova}, I. 2020, arXiv e-prints, arXiv:2012.12378

\bibitem[{{{\"O}stlin} {et~al.}(2001){{\"O}stlin}, {Amram}, {Bergvall}, {Masegosa}, {Boulesteix}, \& {M{\'a}rquez}}]{Ostlin_2001}
{{\"O}stlin}, G., {Amram}, P., {Bergvall}, N., {et~al.} 2001, \aap, 374, 800

\bibitem[{{{\"O}stlin} {et~al.}(2024){{\"O}stlin}, {P{\'e}rez-Gonz{\'a}lez}, {Melinder}, {Gillman}, {Iani}, {Costantin}, {Boogaard}, {Rinaldi}, {Colina}, {N{\o}rgaard-Nielsen}, {Dicken}, {Greve}, {Wright}, {Alonso-Herrero}, {Alvarez-Marquez}, {Annunziatella}, {Bik}, {Bosman}, {Caputi}, {Crespo Gomez}, {Eckart}, {Garcia-Marin}, {Hjorth}, {Ilbert}, {Jermann}, {Kendrew}, {Labiano}, {Langeroodi}, {Le Fevre}, {Libralato}, {Meyer}, {Moutard}, {Peissker}, {Pye}, {Tikkanen}, {Topinka}, {Walter}, {Ward}, {van der Werf}, {van Dishoeck}, {Henning}, {Lagage}, {Ray}, \& {Vandenbussche}}]{Oestlin_MIDIS_2024}
{{\"O}stlin}, G., {P{\'e}rez-Gonz{\'a}lez}, P.~G., {Melinder}, J., {et~al.} 2024, arXiv e-prints, arXiv:2411.19686

\bibitem[{{Pan} {et~al.}(2019){Pan}, {Lin}, {Hsieh}, {Barrera-Ballesteros}, {S{\'a}nchez}, {Hsu}, {Keenan}, {Tissera}, {Boquien}, {Dai}, {Knapen}, {Riffel}, {Argudo-Fern{\'a}ndez}, {Xiao}, \& {Yuan}}]{Pan_2019}
{Pan}, H.-A., {Lin}, L., {Hsieh}, B.-C., {et~al.} 2019, \apj, 881, 119

\bibitem[{pandas~development team(2020)}]{Pandas}
pandas~development team, T. 2020, pandas-dev/pandas: Pandas, vlatest,  Zenodo, doi:10.5281/zenodo.3509134.
\newblock \url{https://doi.org/10.5281/zenodo.3509134}

\bibitem[{{Pearson} {et~al.}(2019){Pearson}, {Wang}, {Alpaslan}, {Baldry}, {Bilicki}, {Brown}, {Grootes}, {Holwerda}, {Kitching}, {Kruk}, \& {van der Tak}}]{Pearson_2019}
{Pearson}, W.~J., {Wang}, L., {Alpaslan}, M., {et~al.} 2019, \aap, 631, A51

\bibitem[{{Peng} {et~al.}(2010){Peng}, {Lilly}, {Kova{\v{c}}}, {Bolzonella}, {Pozzetti}, {Renzini}, {Zamorani}, {Ilbert}, {Knobel}, {Iovino}, {Maier}, {Cucciati}, {Tasca}, {Carollo}, {Silverman}, {Kampczyk}, {de Ravel}, {Sanders}, {Scoville}, {Contini}, {Mainieri}, {Scodeggio}, {Kneib}, {Le F{\`e}vre}, {Bardelli}, {Bongiorno}, {Caputi}, {Coppa}, {de la Torre}, {Franzetti}, {Garilli}, {Lamareille}, {Le Borgne}, {Le Brun}, {Mignoli}, {Perez Montero}, {Pello}, {Ricciardelli}, {Tanaka}, {Tresse}, {Vergani}, {Welikala}, {Zucca}, {Oesch}, {Abbas}, {Barnes}, {Bordoloi}, {Bottini}, {Cappi}, {Cassata}, {Cimatti}, {Fumana}, {Hasinger}, {Koekemoer}, {Leauthaud}, {Maccagni}, {Marinoni}, {McCracken}, {Memeo}, {Meneux}, {Nair}, {Porciani}, {Presotto}, \& {Scaramella}}]{Peng_2010}
{Peng}, Y.-j., {Lilly}, S.~J., {Kova{\v{c}}}, K., {et~al.} 2010, \apj, 721, 193

\bibitem[{{P{\'e}rez-Gonz{\'a}lez} {et~al.}(2023){P{\'e}rez-Gonz{\'a}lez}, {Costantin}, {Langeroodi}, {Rinaldi}, {Annunziatella}, {Ilbert}, {Colina}, {N{\o}rgaard-Nielsen}, {Greve}, {{\"O}stlin}, {Wright}, {Alonso-Herrero}, {{\'A}lvarez-M{\'a}rquez}, {Caputi}, {Eckart}, {Le F{\`e}vre}, {Labiano}, {Garc{\'\i}a-Mar{\'\i}n}, {Hjorth}, {Kendrew}, {Pye}, {Tikkanen}, {van der Werf}, {Walter}, {Ward}, {Bik}, {Boogaard}, {Bosman}, {G{\'o}mez}, {Gillman}, {Iani}, {Jermann}, {Melinder}, {Meyer}, {Moutard}, {van Dishoek}, {Henning}, {Lagage}, {Guedel}, {Peissker}, {Ray}, {Vandenbussche}, {Garc{\'\i}a-Argum{\'a}nez}, \& {Mar{\'\i}a M{\'e}rida}}]{Perez_Gonzalez_2023}
{P{\'e}rez-Gonz{\'a}lez}, P.~G., {Costantin}, L., {Langeroodi}, D., {et~al.} 2023, \apjl, 951, L1

\bibitem[{{Perrin} {et~al.}(2014){Perrin}, {Sivaramakrishnan}, {Lajoie}, {Elliott}, {Pueyo}, {Ravindranath}, \& {Albert}}]{webbpsf}
{Perrin}, M.~D., {Sivaramakrishnan}, A., {Lajoie}, C.-P., {et~al.} 2014, in Society of Photo-Optical Instrumentation Engineers (SPIE) Conference Series, Vol. 9143, Space Telescopes and Instrumentation 2014: Optical, Infrared, and Millimeter Wave, ed. J.~{Oschmann}, Jacobus~M., M.~{Clampin}, G.~G. {Fazio}, \& H.~A. {MacEwen}, 91433X

\bibitem[{{Popesso} {et~al.}(2023){Popesso}, {Concas}, {Cresci}, {Belli}, {Rodighiero}, {Inami}, {Dickinson}, {Ilbert}, {Pannella}, \& {Elbaz}}]{Popesso_2023}
{Popesso}, P., {Concas}, A., {Cresci}, G., {et~al.} 2023, \mnras, 519, 1526

\bibitem[{{Renaud} {et~al.}(2019){Renaud}, {Bournaud}, {Agertz}, {Kraljic}, {Schinnerer}, {Bolatto}, {Daddi}, \& {Hughes}}]{Renaud_2019}
{Renaud}, F., {Bournaud}, F., {Agertz}, O., {et~al.} 2019, \aap, 625, A65

\bibitem[{{Renaud} {et~al.}(2022){Renaud}, {Segovia Otero}, \& {Agertz}}]{Renaud_2022}
{Renaud}, F., {Segovia Otero}, {\'A}., \& {Agertz}, O. 2022, \mnras, 516, 4922

\bibitem[{{Rinaldi} {et~al.}(2022){Rinaldi}, {Caputi}, {van Mierlo}, {Ashby}, {Caminha}, \& {Iani}}]{Rinaldi_2022}
{Rinaldi}, P., {Caputi}, K.~I., {van Mierlo}, S.~E., {et~al.} 2022, \apj, 930, 128

\bibitem[{{Rinaldi} {et~al.}(2023){Rinaldi}, {Caputi}, {Costantin}, {Gillman}, {Iani}, {P{\'e}rez-Gonz{\'a}lez}, {{\"O}stlin}, {Colina}, {Greve}, {Noorgard-Nielsen}, {Wright}, {Alonso-Herrero}, {{\'A}lvarez-M{\'a}rquez}, {Eckart}, {Garc{\'\i}a-Mar{\'\i}n}, {Hjorth}, {Ilbert}, {Kendrew}, {Labiano}, {Le F{\`e}vre}, {Pye}, {Tikkanen}, {Walter}, {van der Werf}, {Ward}, {Annunziatella}, {Azzollini}, {Bik}, {Boogaard}, {Bosman}, {Crespo G{\'o}mez}, {Jermann}, {Langeroodi}, {Melinder}, {Meyer}, {Moutard}, {Peissker}, {Topinka}, {van Dishoeck}, {G{\"u}del}, {Henning}, {Lagage}, {Ray}, {Vandenbussche}, {Waelkens}, {Navarro-Carrera}, \& {Kokorev}}]{Rinaldi_2023}
{Rinaldi}, P., {Caputi}, K.~I., {Costantin}, L., {et~al.} 2023, \apj, 952, 143

\bibitem[{{Rinaldi} {et~al.}(2024){Rinaldi}, {Caputi}, {Iani}, {Costantin}, {Gillman}, {Perez Gonzalez}, {{\"O}stlin}, {Colina}, {Greve}, {N{\o}rgard-Nielsen}, {Wright}, {{\'A}lvarez-M{\'a}rquez}, {Eckart}, {Garc{\'\i}a-Mar{\'\i}n}, {Hjorth}, {Ilbert}, {Kendrew}, {Labiano}, {Le F{\`e}vre}, {Pye}, {Tikkanen}, {Walter}, {van der Werf}, {Ward}, {Annunziatella}, {Azzollini}, {Bik}, {Boogaard}, {Bosman}, {Crespo G{\'o}mez}, {Jermann}, {Langeroodi}, {Melinder}, {Meyer}, {Moutard}, {Peissker}, {van Dishoeck}, {G{\"u}del}, {Henning}, {Lagage}, {Ray}, {Vandenbussche}, {Waelkens}, \& {Dayal}}]{Rinaldi_midis_2024}
{Rinaldi}, P., {Caputi}, K.~I., {Iani}, E., {et~al.} 2024, \apj, 969, 12

\bibitem[{{Rodighiero} {et~al.}(2011){Rodighiero}, {Daddi}, {Baronchelli}, {Cimatti}, {Renzini}, {Aussel}, {Popesso}, {Lutz}, {Andreani}, {Berta}, {Cava}, {Elbaz}, {Feltre}, {Fontana}, {F{\"o}rster Schreiber}, {Franceschini}, {Genzel}, {Grazian}, {Gruppioni}, {Ilbert}, {Le Floch}, {Magdis}, {Magliocchetti}, {Magnelli}, {Maiolino}, {McCracken}, {Nordon}, {Poglitsch}, {Santini}, {Pozzi}, {Riguccini}, {Tacconi}, {Wuyts}, \& {Zamorani}}]{Rodighiero_2011}
{Rodighiero}, G., {Daddi}, E., {Baronchelli}, I., {et~al.} 2011, \apjl, 739, L40

\bibitem[{{Rodr{\'\i}guez Montero} {et~al.}(2019){Rodr{\'\i}guez Montero}, {Dav{\'e}}, {Wild}, {Angl{\'e}s-Alc{\'a}zar}, \& {Narayanan}}]{Montero_2019}
{Rodr{\'\i}guez Montero}, F., {Dav{\'e}}, R., {Wild}, V., {Angl{\'e}s-Alc{\'a}zar}, D., \& {Narayanan}, D. 2019, \mnras, 490, 2139

\bibitem[{{Romeo} \& {Fathi}(2016)}]{Romeo_2016}
{Romeo}, A.~B., \& {Fathi}, K. 2016, \mnras, 460, 2360

\bibitem[{{Rosani} {et~al.}(2020){Rosani}, {Caminha}, {Caputi}, \& {Deshmukh}}]{Rosani_2020}
{Rosani}, G., {Caminha}, G.~B., {Caputi}, K.~I., \& {Deshmukh}, S. 2020, \aap, 633, A159

\bibitem[{{Salmon} {et~al.}(2015){Salmon}, {Papovich}, {Finkelstein}, {Tilvi}, {Finlator}, {Behroozi}, {Dahlen}, {Dav{\'e}}, {Dekel}, {Dickinson}, {Ferguson}, {Giavalisco}, {Long}, {Lu}, {Mobasher}, {Reddy}, {Somerville}, \& {Wechsler}}]{Salmon_2015}
{Salmon}, B., {Papovich}, C., {Finkelstein}, S.~L., {et~al.} 2015, \apj, 799, 183

\bibitem[{{Salpeter}(1955)}]{Salpeter_1955}
{Salpeter}, E.~E. 1955, \apj, 121, 161

\bibitem[{{S{\'a}nchez Almeida} {et~al.}(2014){S{\'a}nchez Almeida}, {Elmegreen}, {Mu{\~n}oz-Tu{\~n}{\'o}n}, \& {Elmegreen}}]{Almeida_2014}
{S{\'a}nchez Almeida}, J., {Elmegreen}, B.~G., {Mu{\~n}oz-Tu{\~n}{\'o}n}, C., \& {Elmegreen}, D.~M. 2014, \aapr, 22, 71

\bibitem[{{Santini} {et~al.}(2017){Santini}, {Fontana}, {Castellano}, {Di Criscienzo}, {Merlin}, {Amorin}, {Cullen}, {Daddi}, {Dickinson}, {Dunlop}, {Grazian}, {Lamastra}, {McLure}, {Micha{\l}owski}, {Pentericci}, \& {Shu}}]{Santini_2017}
{Santini}, P., {Fontana}, A., {Castellano}, M., {et~al.} 2017, \apj, 847, 76

\bibitem[{{Sargent} {et~al.}(2012){Sargent}, {B{\'e}thermin}, {Daddi}, \& {Elbaz}}]{Sargent_2012}
{Sargent}, M.~T., {B{\'e}thermin}, M., {Daddi}, E., \& {Elbaz}, D. 2012, \apjl, 747, L31

\bibitem[{{Schreiber} {et~al.}(2015){Schreiber}, {Pannella}, {Elbaz}, {B{\'e}thermin}, {Inami}, {Dickinson}, {Magnelli}, {Wang}, {Aussel}, {Daddi}, {Juneau}, {Shu}, {Sargent}, {Buat}, {Faber}, {Ferguson}, {Giavalisco}, {Koekemoer}, {Magdis}, {Morrison}, {Papovich}, {Santini}, \& {Scott}}]{Schreiber_2015}
{Schreiber}, C., {Pannella}, M., {Elbaz}, D., {et~al.} 2015, \aap, 575, A74

\bibitem[{{Scoville} {et~al.}(2007){Scoville}, {Aussel}, {Brusa}, {Capak}, {Carollo}, {Elvis}, {Giavalisco}, {Guzzo}, {Hasinger}, {Impey}, {Kneib}, {LeFevre}, {Lilly}, {Mobasher}, {Renzini}, {Rich}, {Sanders}, {Schinnerer}, {Schminovich}, {Shopbell}, {Taniguchi}, \& {Tyson}}]{Scoville_2007}
{Scoville}, N., {Aussel}, H., {Brusa}, M., {et~al.} 2007, \apjs, 172, 1

\bibitem[{{Smit} {et~al.}(2016){Smit}, {Bouwens}, {Labb{\'e}}, {Franx}, {Wilkins}, \& {Oesch}}]{Smit_2016}
{Smit}, R., {Bouwens}, R.~J., {Labb{\'e}}, I., {et~al.} 2016, \apj, 833, 254

\bibitem[{{Sonnett} {et~al.}(2013){Sonnett}, {Meech}, {Jedicke}, {Bus}, {Tonry}, \& {Hainaut}}]{Sonnett_2013}
{Sonnett}, S., {Meech}, K., {Jedicke}, R., {et~al.} 2013, \pasp, 125, 456

\bibitem[{{Sparre} {et~al.}(2015){Sparre}, {Hayward}, {Springel}, {Vogelsberger}, {Genel}, {Torrey}, {Nelson}, {Sijacki}, \& {Hernquist}}]{Sparre_2015}
{Sparre}, M., {Hayward}, C.~C., {Springel}, V., {et~al.} 2015, \mnras, 447, 3548

\bibitem[{{Speagle} {et~al.}(2014){Speagle}, {Steinhardt}, {Capak}, \& {Silverman}}]{Speagle_2014}
{Speagle}, J.~S., {Steinhardt}, C.~L., {Capak}, P.~L., \& {Silverman}, J.~D. 2014, \apjs, 214, 15

\bibitem[{{Stefanon} {et~al.}(2019){Stefanon}, {Labb{\'e}}, {Bouwens}, {Oesch}, {Ashby}, {Caputi}, {Franx}, {Fynbo}, {Illingworth}, {Le F{\`e}vre}, {Marchesini}, {McCracken}, {Milvang-Jensen}, {Muzzin}, \& {van Dokkum}}]{Stefanon_2019}
{Stefanon}, M., {Labb{\'e}}, I., {Bouwens}, R.~J., {et~al.} 2019, \apj, 883, 99

\bibitem[{{Tacchella} {et~al.}(2016){Tacchella}, {Dekel}, {Carollo}, {Ceverino}, {DeGraf}, {Lapiner}, {Mandelker}, \& {Primack Joel}}]{Tacchella_2016}
{Tacchella}, S., {Dekel}, A., {Carollo}, C.~M., {et~al.} 2016, \mnras, 457, 2790

\bibitem[{{Tadaki} {et~al.}(2018){Tadaki}, {Iono}, {Yun}, {Aretxaga}, {Hatsukade}, {Hughes}, {Ikarashi}, {Izumi}, {Kawabe}, {Kohno}, {Lee}, {Matsuda}, {Nakanishi}, {Saito}, {Tamura}, {Ueda}, {Umehata}, {Wilson}, {Michiyama}, {Ando}, \& {Kamieneski}}]{Tadaki_2018}
{Tadaki}, K., {Iono}, D., {Yun}, M.~S., {et~al.} 2018, \nat, 560, 613

\bibitem[{{Taniguchi} {et~al.}(2007){Taniguchi}, {Scoville}, {Murayama}, {Sanders}, {Mobasher}, {Aussel}, {Capak}, {Ajiki}, {Miyazaki}, {Komiyama}, {Shioya}, {Nagao}, {Sasaki}, {Koda}, {Carilli}, {Giavalisco}, {Guzzo}, {Hasinger}, {Impey}, {LeFevre}, {Lilly}, {Renzini}, {Rich}, {Schinnerer}, {Shopbell}, {Kaifu}, {Karoji}, {Arimoto}, {Okamura}, \& {Ohta}}]{Taniguchi_2007}
{Taniguchi}, Y., {Scoville}, N., {Murayama}, T., {et~al.} 2007, \apjs, 172, 9

\bibitem[{{Taylor}(2005)}]{Topcat}
{Taylor}, M.~B. 2005, in Astronomical Society of the Pacific Conference Series, Vol. 347, Astronomical Data Analysis Software and Systems XIV, ed. P.~{Shopbell}, M.~{Britton}, \& R.~{Ebert}, 29

\bibitem[{{van Mierlo} {et~al.}(2022){van Mierlo}, {Caputi}, {Ashby}, {Atek}, {Bolzonella}, {Bowler}, {Brammer}, {Conselice}, {Cuby}, {Dayal}, {D{\'\i}az-S{\'a}nchez}, {Finkelstein}, {Hoekstra}, {Humphrey}, {Ilbert}, {McCracken}, {Milvang-Jensen}, {Oesch}, {Pello}, {Rodighiero}, {Schirmer}, {Toft}, {Weaver}, {Wilkins}, {Willott}, {Zamorani}, {Amara}, {Auricchio}, {Baldi}, {Bender}, {Bodendorf}, {Bonino}, {Branchini}, {Brescia}, {Brinchmann}, {Camera}, {Capobianco}, {Carbone}, {Carretero}, {Castellano}, {Cavuoti}, {Cimatti}, {Cledassou}, {Congedo}, {Conversi}, {Copin}, {Corcione}, {Courbin}, {Da Silva}, {Degaudenzi}, {Douspis}, {Dubath}, {Dupac}, {Dusini}, {Farrens}, {Ferriol}, {Frailis}, {Franceschi}, {Franzetti}, {Fumana}, {Galeotta}, {Garilli}, {Gillard}, {Gillis}, {Giocoli}, {Grazian}, {Grupp}, {Haugan}, {Holmes}, {Hormuth}, {Hornstrup}, {Jahnke}, {K{\"u}mmel}, {Kiessling}, {Kilbinger}, {Kitching}, {Kohley}, {Kunz}, {Kurki-Suonio}, {Laureijs}, {Ligori}, {Lilje}, {Lloro}, {Maiorano}, {Mansutti}, {Marggraf},
  {Markovic}, {Marulli}, {Massey}, {Maurogordato}, {Medinaceli}, {Meneghetti}, {Merlin}, {Meylan}, {Moresco}, {Moscardini}, {Munari}, {Niemi}, {Padilla}, {Paltani}, {Pasian}, {Pedersen}, {Pettorino}, {Pires}, {Poncet}, {Popa}, {Pozzetti}, {Raison}, {Renzi}, {Rhodes}, {Riccio}, {Romelli}, {Rossetti}, {Saglia}, {Sapone}, {Sartoris}, {Schneider}, {Secroun}, {Sirignano}, {Sirri}, {Stanco}, {Starck}, {Surace}, {Tallada-Cresp{\'\i}}, {Taylor}, {Tereno}, {Toledo-Moreo}, {Torradeflot}, {Tutusaus}, {Valentijn}, {Valenziano}, {Vassallo}, {Wang}, {Zacchei}, {Zoubian}, {Andreon}, {Bardelli}, {Boucaud}, {Graci{\'a}-Carpio}, {Maino}, {Mauri}, {Mei}, {Sureau}, {Zucca}, {Aussel}, {Baccigalupi}, {Balaguera-Antol{\'\i}nez}, {Biviano}, {Blanchard}, {Borgani}, {Bozzo}, {Burigana}, {Cabanac}, {Calura}, {Cappi}, {Carvalho}, {Casas}, {Castignani}, {Colodro-Conde}, {Cooray}, {Coupon}, {Courtois}, {Crocce}, {Cucciati}, {Davini}, {Dole}, {Escartin}, {Escoffier}, {Fabricius}, {Farina}, {Ganga}, {Garc{\'\i}a-Bellido}, {George},
  {Giacomini}, {Gozaliasl}, {Gwyn}, {Hook}, {Huertas-Company}, {Kansal}, {Kashlinsky}, {Keihanen}, {Kirkpatrick}, {Lindholm}, {Maoli}, {Martinelli}, {Martinet}, {Maturi}, {Metcalf}, {Monaco}, {Morgante}, {Nucita}, {Patrizii}, {Peel}, {Pollack}, {Popa}, {Porciani}, {Potter}, {Reimberg}, {S{\'a}nchez}, {Scottez}, {Sefusatti}, {Stadel}, {Teyssier}, {Valiviita}, \& {Viel}}]{van_Mierlo_2022}
{van Mierlo}, S.~E., {Caputi}, K.~I., {Ashby}, M., {et~al.} 2022, \aap, 666, A200

\bibitem[{{Ventou} {et~al.}(2019){Ventou}, {Contini}, {Bouch{\'e}}, {Epinat}, {Brinchmann}, {Inami}, {Richard}, {Schroetter}, {Soucail}, {Steinmetz}, \& {Weilbacher}}]{Ventou_2019}
{Ventou}, E., {Contini}, T., {Bouch{\'e}}, N., {et~al.} 2019, \aap, 631, A87

\bibitem[{Virtanen {et~al.}(2020)Virtanen, Gommers, Oliphant, Haberland, Reddy, Cournapeau, Burovski, Peterson, Weckesser, Bright, {van der Walt}, Brett, Wilson, Millman, Mayorov, Nelson, Jones, Kern, Larson, Carey, Polat, Feng, Moore, {VanderPlas}, Laxalde, Perktold, Cimrman, Henriksen, Quintero, Harris, Archibald, Ribeiro, Pedregosa, {van Mulbregt}, \& {SciPy 1.0 Contributors}}]{Scipy}
Virtanen, P., Gommers, R., Oliphant, T.~E., {et~al.} 2020, Nature Methods, 17, 261

\bibitem[{{Wang} {et~al.}(2019){Wang}, {Lilly}, {Pezzulli}, \& {Matthee}}]{Wang_2019}
{Wang}, E., {Lilly}, S.~J., {Pezzulli}, G., \& {Matthee}, J. 2019, \apj, 877, 132

\bibitem[{{Whitaker} {et~al.}(2012){Whitaker}, {van Dokkum}, {Brammer}, \& {Franx}}]{Whitaker_2012}
{Whitaker}, K.~E., {van Dokkum}, P.~G., {Brammer}, G., \& {Franx}, M. 2012, \apjl, 754, L29

\bibitem[{{Whitaker} {et~al.}(2014){Whitaker}, {Franx}, {Leja}, {van Dokkum}, {Henry}, {Skelton}, {Fumagalli}, {Momcheva}, {Brammer}, {Labb{\'e}}, {Nelson}, \& {Rigby}}]{Whitaker_2014}
{Whitaker}, K.~E., {Franx}, M., {Leja}, J., {et~al.} 2014, \apj, 795, 104

\bibitem[{{Whitaker} {et~al.}(2019){Whitaker}, {Ashas}, {Illingworth}, {Magee}, {Leja}, {Oesch}, {van Dokkum}, {Mowla}, {Bouwens}, {Franx}, {Holden}, {Labb{\'e}}, {Rafelski}, {Teplitz}, \& {Gonzalez}}]{Whitaker_2019}
{Whitaker}, K.~E., {Ashas}, M., {Illingworth}, G., {et~al.} 2019, \apjs, 244, 16

\bibitem[{{Williams} {et~al.}(2023){Williams}, {Tacchella}, {Maseda}, {Robertson}, {Johnson}, {Willott}, {Eisenstein}, {Willmer}, {Ji}, {Hainline}, {Helton}, {Alberts}, {Baum}, {Bhatawdekar}, {Boyett}, {Bunker}, {Carniani}, {Charlot}, {Chevallard}, {Curtis-Lake}, {de Graaff}, {Egami}, {Franx}, {Kumari}, {Maiolino}, {Nelson}, {Rieke}, {Sandles}, {Shivaei}, {Simmonds}, {Smit}, {Suess}, {Sun}, {{\"U}bler}, \& {Witstok}}]{Williams_2023}
{Williams}, C.~C., {Tacchella}, S., {Maseda}, M.~V., {et~al.} 2023, \apjs, 268, 64

\bibitem[{{Zhang} {et~al.}(2023){Zhang}, {Li}, {Leja}, {Whitaker}, {Nersesian}, {Bezanson}, \& {van der Wel}}]{Zhang_2023}
{Zhang}, J., {Li}, Y., {Leja}, J., {et~al.} 2023, \apj, 952, 6

\end{thebibliography}
\bibliographystyle{aasjournal}
\end{document}